\documentclass[acmsmall]{acmart}

\newcommand{\toolname}{\textsc{SafeEdit}}

\usepackage{enumerate}
\usepackage{ragged2e}
\usepackage{pifont}
\usepackage{stfloats}
\usepackage{amsmath,amsfonts}
\usepackage{graphicx}
\usepackage{textcomp}
\usepackage{xcolor}
\usepackage[normalem]{ulem} 
\usepackage{balance}
\usepackage{hyperref}
\newtheorem{definition}{Definition}
\usepackage{url}
\usepackage{tabularx}  

\usepackage{wrapfig}
\usepackage{rotating}

\usepackage{multirow}
\usepackage[utf8]{inputenc}
\usepackage{graphicx}
\usepackage{textcomp}
\usepackage{color,xcolor}
\usepackage{url}
\usepackage{graphicx}
\usepackage{stmaryrd}
\usepackage{verbatimbox}
\usepackage{enumerate}
\usepackage[shortlabels]{enumitem}
\usepackage{bm}
\usepackage{soul} 
\usepackage{makecell}
\usepackage{booktabs}

\usepackage{multirow}

\usepackage{enumitem}
\usepackage{colortbl}
\usepackage{subcaption}  

\definecolor{bestbg}{gray}{0.9} 

\definecolor{negred}{RGB}{180, 0, 0}

\newcommand{\cbg}{\cellcolor{bestbg}}

\newcommand{\best}[1]{\cellcolor{gray!20}\textbf{#1}}
\newcommand{\second}[1]{\underline{#1}}
\usepackage{pifont} 

\usepackage{listings}
\usepackage[most]{tcolorbox}
\tcbuselibrary{listingsutf8,breakable}
\definecolor{lsframe}{HTML}{CCCCCC}   
\definecolor{lsbg}{HTML}{F7F7F7}      

\newcommand{\find}[2]{
\begin{tcolorbox}[toprule=0mm,bottomrule=0mm,left=1pt,right=2pt,top=2pt,bottom=2pt,breakable]%
\em #1
\end{tcolorbox}
}

\lstdefinestyle{paperlisting}{
  basicstyle=\ttfamily\scriptsize,
  keywordstyle=\color{blue}\bfseries,
  commentstyle=\itshape\color{teal!70!black},
  stringstyle=\color{magenta!70!black},
  frame=lines,                        
  framerule=0.4pt,
  rulecolor=\color{lsframe},
  backgroundcolor=\color{lsbg},
  xleftmargin=0.6em, xrightmargin=0.6em,
  aboveskip=0.5\baselineskip, belowskip=0.5\baselineskip,
  showstringspaces=false,
  breaklines=true, columns=fullflexible, tabsize=4,
  captionpos=t                        
}

\newcommand{\myfinding}[2]{
\begin{center}
\begin{tcolorbox}[colback=gray!15, colframe=black, boxsep= -0.15cm, middle=-0.15cm, breakable]
\textbf{Answer to RQ{#1}:}
{#2}
\end{tcolorbox}
\end{center}
}

\usepackage[ruled,linesnumbered,noend]{algorithm2e}
\SetKwProg{Fn}{Function}{}{}
\SetKwFunction{InitializeLangPrior}{InitializeLangPrior}
\SetKwFunction{SelectLanguage}{SelectLanguage}
\SetKwFunction{UpdateStats}{UpdateStats}
\SetKwFunction{InitializeSeedBank}{InitializeSeedBank}
\SetKwFunction{InitSurvivorPools}{InitSurvivorPools}
\SetKwFunction{SampleSeed}{SampleSeed}
\SetKwFunction{UpdateSurvivorPool}{UpdateSurvivorPool}
\SetKwFunction{SoftmaxSample}{SoftmaxSample}
\DontPrintSemicolon

\usepackage{tcolorbox}
\tcbuselibrary{breakable,skins}

\makeatletter
\newtheoremstyle{plainnoparen}
  {6pt} {6pt}
  {\itshape}
  {}
  {\bfseries}
  {.}
  {.5em}
  {\thmname{#1}~\thmnumber{#2}.\ \thmnote{\itshape #3}} 
\makeatother

\theoremstyle{plainnoparen}

\usepackage{listings}
\tcbuselibrary{listings, skins, breakable}

\newtcblisting{mycode}[2][]{
  listing engine=listings,
  listing options={
    language=C,
    basicstyle=\ttfamily\scriptsize,
    numbers=left,
    numberstyle=\tiny,
    numbersep=6pt,
    breaklines=true,
    showstringspaces=false,
    columns=fullflexible,
    keepspaces=true
  },
  listing only,
  colback=green!3!white,
  colframe=green!40!black,
  arc=4pt,
  boxrule=0.5pt,
  top=0mm,
  bottom=0mm,
  toptitle=0mm,
  bottomtitle=0mm,
  left=6mm,
  width=\linewidth,
  title=#2,
  #1
}

\usepackage{xspace}
\newcommand{\ie}{\textit{i.e.,}\xspace}
\newcommand{\eg}{\textit{e.g.,}\xspace}

\newcommand{\etal}{\textit{et al.}\xspace}

\usepackage{fontawesome} 
\usepackage{microtype}
\usepackage{minted}
\counterwithin{listing}{section}  

\AtBeginDocument{%
  }

\setcopyright{cc}
\setcctype{by}
\acmDOI{10.1145/3832254}
\acmYear{2026}
\acmJournal{PACMSE}
\acmVolume{3}
\acmNumber{ISSTA}
\acmArticle{ISSTA163}
\acmMonth{10}
\acmSubmissionID{issta26main-p1843-p}
\received{2026-01-30}
\received[accepted]{2026-06-25}

\begin{document}

\title{Understanding and Improving Model Editing for Secure Code Generation}

\author{Weifeng Sun}
\orcid{0000-0001-6013-1369}
\affiliation{%
  \institution{Singapore Management University}
  \city{Singapore}
  \country{Singapore}
}
\email{wfsun@smu.edu.sg}

\author{Quanjun Zhang}
\authornote{corresponding author.}
\orcid{0000-0002-2495-3805}
\affiliation{%
  \institution{Nanjing University of Science and Technology}
  \city{Nanjing}
  \country{China}
}
\email{quanjunzhang@njust.edu.cn}

\author{Yuchen Chen}
\orcid{0000-0002-3380-5564}
\affiliation{%
  \institution{Nanjing University}
  \city{Nanjing}
  \country{China}
}
\email{yuc.chen@smail.nju.edu.cn}

\author{Chengran Yang}
\orcid{0000-0001-6100-8127}
\affiliation{%
  \institution{Singapore Management University}
  \city{Singapore}
  \country{Singapore}
}
\email{cryang@smu.edu.sg}

\author{Gou Tan}
\orcid{0009-0008-6580-1470}
\affiliation{%
  \institution{Sun Yat-sen University}
  \city{Guangzhou}
  \country{China}
}
\email{tang29@mail2.sysu.edu.cn}

\author{David Lo}
\orcid{0000-0002-4367-7201}
\affiliation{%
  \institution{Singapore Management University}
  \city{Singapore}
  \country{Singapore}
}
\email{davidlo@smu.edu.sg}

\begin{abstract}

Large language models (LLMs) are widely used for code generation, yet they can reproduce vulnerable code implementations learned from insecure patterns in training data.
Prior work has primarily explored \emph{inference-time hardening} to reduce insecure generations without updating the target model.
While effective, this paradigm couples security behavior to the auxiliary component and incurs additional runtime overhead.
This paper presents the first systematic empirical study of applying \emph{model editing} as the \emph{model-level} hardening mechanism for secure code generation.
Unlike inference-time interventions, model editing updates a small subset of parameters to inject security-relevant knowledge directly into the target LLM.
We evaluate 3 state-of-the-art editing methods across diverse LLM families and compare them with CoSec, a representative inference-time hardening approach, focusing on:
(i) security effectiveness and robustness,
(ii) generalization to unseen vulnerabilities, and 
(iii) functional correctness on general programming tasks.
Our results show that model editing yields substantially larger security gains than CoSec on seen vulnerability types, improving security ratios by 15\%–25\% over vanilla models, and these gains remain stable under prompt perturbations. 
However, security improvements do not always transfer reliably to unseen vulnerabilities and induce functional regressions on general programming tasks, even for UltraEdit, the best-performing editing method in our evaluation.
To mitigate this side effect, we propose \toolname{}, a post-edit refinement method that combines functional tuning with edit-aware regularization.
Across eight target LLMs, \toolname{} improves Pass@1 over UltraEdit by +11.73/+13.70/+15.50 percentage points at $T=0.1/0.4/0.8$, while largely preserving security.
Compared with CoSec, it achieves relative security-ratio gains of +7.54\%--12.04\%.
Additional evaluation on CodeGuard+ confirms that \toolname{} improves joint secure-and-correct generation.
Importantly, \toolname{} and CoSec are complementary: combining them can improve security under stochastic decoding while maintaining strong functional correctness.  
Finally, we analyze efficiency and key design factors, showing that model editing is more efficient than CoSec and sensitive to editing depth, parameter location, and injected context.
Overall, our work provides evidence-backed guidance for applying model editing to secure code generation.

\end{abstract}



\maketitle
\vspace{-3mm}
\section{Introduction}
\label{sec:intro}
The rapid progress of large language models (LLMs), \eg GPT-3~\cite{Brown2020}, LLaMA~\cite{Touvron2023}, and GLM~\cite{Du2022}, has reshaped software practices~\cite{zhang2026survey,Xu2022}.
Their ability to understand natural language and generate code facilitates various code-related tasks~\cite{ li2026retrieval, li2025retrieval, sun2023revisiting, fan2025exploring}, such as code translation~\cite{Yang2024,Tehrani2024, Sun2025}, code completion~\cite{Izadi2024, Wang2024}, test generation~\cite{zhang2025large}, and program repair~\cite{Silva2025,zhang2024systematic,zhang2023gamma}.
Consequently, LLM-powered assistants, including GitHub Copilot~\cite{GitHubCopilot2022} and CodeGeeX~\cite{ZAI2024CodeGeeX},  have been widely adopted in development environments.
However, prior studies show that LLMs can generate insecure code~\cite{Khoury2023,Majdinasab2024,Pearce2025}, since training corpora contain vulnerable implementations that can be reproduced at inference time~\cite{Chen2024a,Dou2024,Ji2023}.
Therefore, mitigating insecure code generation has become a pressing practical concern.

To address this challenge, previous work has explored \emph{\textbf{inference-time hardening}} strategies that modify decoding behavior without updating the target model parameters, with the aim of minimizing insecure code generation while preserving \textit{functional correctness}, \ie ensuring that the generated code still implements the intended requirements.
A representative approach is CoSec~\cite{Li2024, Li2025}, which steers decoding by incorporating an auxiliary security model to bias token selection toward safer continuations.
Although effective, such designs introduce two inherent limitations:
1) their effectiveness is closely tied to the capacity and generalization of the auxiliary model, and 
2) they introduce additional runtime overhead due to multi-model decoding, which is non-negligible in latency-sensitive and resource-constrained environments.

A natural alternative is \emph{\textbf{model-level updating}}, \eg retraining or fine-tuning the target LLM, to improve security behavior without relying on auxiliary models at inference time, such as SVEN~\cite{He2023} and SafeCoder~\cite{He2024}.
However, such methods are computationally expensive, and previous work~\cite{Peng2025} has shown that their broad parameter updates significantly disturb general coding behavior, causing regressions on unrelated tasks.
Notably, \textbf{\emph{model editing}} has been proposed as a targeted form of model-level update~\cite{Yao2023}, performing localized parameter changes to enforce specific behavioral updates while largely preserving the original capabilities of the model.


Model editing has recently attracted increasing attention, but its potential for secure code generation remains largely unexplored.
Existing studies apply model editing to token-level code correction~\cite{Gu2023} or API-update-oriented code tasks~\cite{Li2024a}, without considering vulnerability-oriented scenarios or security criteria.
Thus, it remains unclear \textit{whether model editing can effectively strengthen the security of LLM-generated code}.
To fill this gap, we conduct the first systematic empirical study of model editing for secure code generation, examining its feasibility, effectiveness, limitations, and comparison with CoSec, a SOTA inference-time hardening approach.
We attempt to address the following research questions:

\vspace{-0.1cm}
\begin{itemize}[leftmargin=1em]
\item \textbf{Security Effectiveness.}
We assess whether model editing improves secure code generation under vulnerability-relevant contexts, relative to CoSec, and whether these improvements are robust to prompt perturbations.
\textbf{Results}:
Model editing achieves substantially higher security improvements than CoSec, often improving the security ratio by 15--25\% over vanilla (original) models.
Moreover, these improvements remain largely stable under prompt
perturbations.
In contrast, CoSec provides limited improvements and even reduces security when the auxiliary model is much smaller than the target LLM.

\item \textbf{Security Generalization.}
We evaluate how well security improvements from model editing transfer to unseen vulnerability types.
\textbf{Results:}
Strong security improvements on seen vulnerabilities do not necessarily translate to good generalization.
Most editing methods retain part of their improvements on unseen CWE categories, whereas CoSec typically stays close to (or slightly above) the vanilla model on unseen cases.

\item \textbf{Functional Correctness.}
We evaluate how security hardening affects the functional correctness of base models.
\textbf{Results:}
Both model editing methods and CoSec introduce non-negligible degradation in functional correctness.
However, model editing tends to incur larger regressions, especially under cumulative edits.
In comparison, CoSec achieves limited security improvements but better preserves functional correctness under high-temperature decoding conditions.

\item \textbf{Enhanced Approach.}
Motivated by our empirical findings, we propose \toolname{}, a post-edit refinement method.
\toolname{} combines functional tuning with edit-aware regularization to mitigate functional regressions while preserving the injected security behavior.
\textbf{Results:}
Compared to UltraEdit, which achieves the most favorable security–correctness trade-off among the evaluated editing methods, \toolname{} restores functional correctness without sacrificing security effectiveness.
Compared to CoSec, \toolname{} improves both security (up to 26.04\% security ratio) and correctness on average.
We further find that \toolname{} and CoSec offer complementary advantages when combined.
\end{itemize}




Furthermore, we conduct extended experiments to provide additional discussions about secure code generation via model editing from different aspects:

\begin{itemize}[leftmargin=1em]
\item \textbf{Efficiency.}
We compare training overhead and inference latency.
\textbf{Results:}
Model editing is consistently more efficient than CoSec: UltraEdit completes editing within 50--100 seconds even for 8B-scale models, whereas CoSec requires hundreds to thousands of seconds to train an auxiliary security model and increases inference latency by 2--3$\times$.

\item \textbf{Design Choices.}
We ablate editing depth, editing dataset size, parameter location, and injected context.
\textbf{Results:}
Editing performance is sensitive to these factors; overly shallow or aggressive edits may improve security but often cause larger correctness degradation.
\vspace{-0.1cm}
\end{itemize}

Overall, model editing is not a plug-and-play solution: careful design choices are needed to avoid overfitting to security cues at the expense of general code generation.


\textbf{Novelty \& Contributions.}
 The main contributions of this paper are as follows:
 
\begin{itemize}[left=0pt]
\item[(1)]
\vspace{-0.1cm}
\textit{\textbf{New Empirical Perspective.}}
We present the first systematic empirical study on applying \emph{model editing} to secure code generation, investigating whether directly modifying LLM parameters can mitigate insecure code generation and how such approaches compare with CoSec, the state-of-the-art method of secure code generation.
\item[(2)]
\textit{\textbf{Comprehensive Evaluation.}}
We evaluate multiple LLM families and hardening methods across security effectiveness, functional correctness, generalization, robustness, and efficiency, revealing a clear security--correctness trade-off.

\item[(3)]
\textit{\textbf{Effective Technique.}}
We propose \toolname{}, a post-edit refinement method that combines functional tuning with edit-aware regularization to improve functional correctness while retaining security gains.

\item[(4)]
\textit{\textbf{Design Insights for Security Editing.}}
Through targeted ablation studies, we analyze how editing depth, editing dataset size, parameter location, and injected context size affect security and correctness, providing concrete insights into the design of effective model editing strategies.


\end{itemize}

\vspace{-0.3cm}
\section{Study Design}

\begin{figure}
    \centering
    \includegraphics[width=0.95\linewidth]{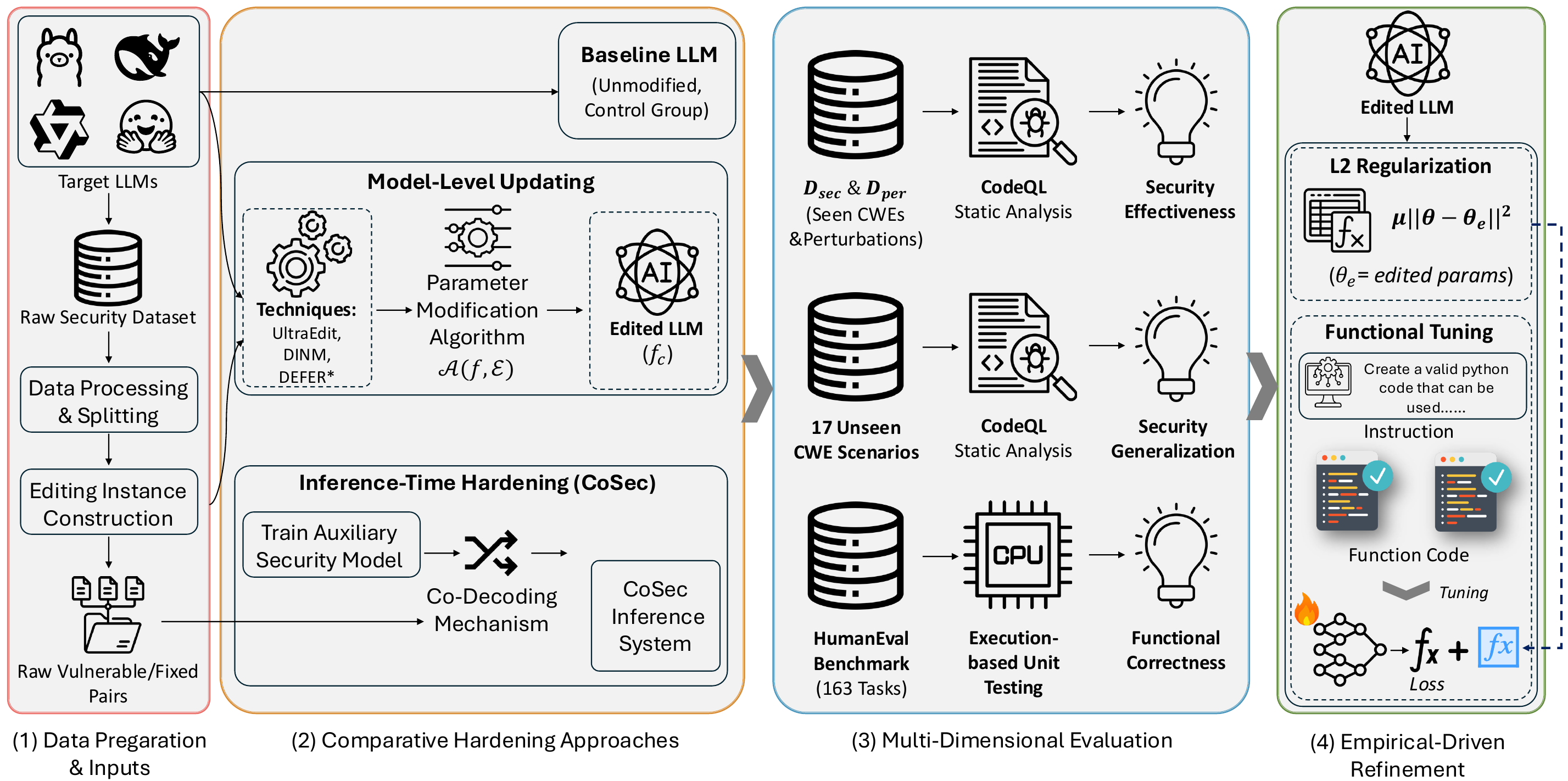}
    \vspace{-0.4cm}
    \caption{Overview of our experimental pipeline for secure code generation hardening, comparing model editing with inference-time hardening (CoSec) across security, generalization, functional correctness.}
    \label{fig:workflow}
\end{figure}

\subsection{Overall Framework}

Figure~\ref{fig:workflow} illustrates our experimental pipeline.
We first construct security editing instances covering multiple CWE categories, and apply model editing to obtain edited LLMs via localized parameter updates.
We compare model editing with CoSec, a representative inference-time hardening baseline.
We then evaluate each method from three perspectives: (i) \emph{security effectiveness} on seen CWEs, including robustness to prompt perturbations; (ii) \emph{security generalization} on unseen CWEs with CodeQL-based static analysis; and (iii) \emph{functional correctness} on HumanEval with execution-based testing.
Finally, motivated by the observed security--correctness trade-off, we propose a post-edit refinement method that combines functional tuning with edit-aware regularization to recover functional correctness while preserving injected security behavior.

\vspace{-0.3cm}
\subsection{Task Definition}
\label{sec:task_def}

\find{
\begin{definition}[Security Editing Instance]
\label{def:security_edit_instance}
Let $f$ denote a target LLM.
A security-driven editing instance is defined as $M_i = (X_i, \hat{Y}_i, Y_i, \text{CWE}_i)$,
where (i) $X_i$ is a vulnerability-relevant code prefix;
(ii) $\hat{Y}_i = f(X_i)$ is the completion produced by $f$ (reflecting the pre-editing behavior);
(iii) $Y_i$ is a secure suffix (or completion) that removes the $\text{CWE}_i$ under the same context $X_i$; and
(iv) $\text{CWE}_i$ denotes the vulnerability category associated with $M_i$.
\end{definition}
}

\find{
\begin{definition}[Security Editing Dataset]
\label{def:security_edit_dataset}
A security editing dataset is a collection of editing instances $\mathcal{E} = \{M_1, M_2, \ldots, M_n\}$,
covering multiple CWEs, where each $M_i$ follows Definition~\ref{def:security_edit_instance}.
\end{definition}
}

\find{
\begin{definition}[Security Hardening via Model Editing]
\label{def:security_hardening_editing}
Given a target model $f$ and a security editing dataset $\mathcal{E}$, a model editing algorithm $\mathcal{A}$ produces an edited model $f_e = \mathcal{A}(f, \mathcal{E})$
where the objective is to update the model such that it generates secure code for vulnerability-relevant inputs while preserving its general coding capability.
\end{definition}
}

Let $\mathcal{T}_{seen}$ denote all CWE categories in $\mathcal{E}$ and $\mathcal{T}_{unseen}$ be the remaining CWE categories in evaluation, where $\mathcal{T}_{seen}\cap\mathcal{T}_{unseen}=\emptyset$.
Let $\mathcal{U}$ denote a set of representative programming problems for functional evaluation.
The edited model $f_e$ is assessed along the following evaluation objectives:

\find{
\begin{definition}[Evaluation Objectives]
\label{def:post_edit_objectives}

\smallskip
\noindent\textbf{(1) Security Effectiveness.}
$f_e$ improves secure code generation on vulnerability-relevant inputs whose CWE belongs to $\mathcal{T}_{seen}$, including but not limited to instances in $\mathcal{E}$.
\smallskip
\noindent\textbf{(2) Security Generalization.}
$f_e$ improves secure code generation on inputs whose CWE belongs to $\mathcal{T}_{unseen}$.
\smallskip
\noindent\textbf{(3) Functional Correctness.}
$f_e$ preserves general coding capability on standard programming tasks:
$f_e(U) \approx f(U)$, $\forall U \in \mathcal{U}$.
\end{definition}
}

Following established practice in model editing research~\cite{Meng2022, Zhang2024a, Zhu2020},
we apply each editing technique to the full editing set $\mathcal{E}$ regardless of whether the target LLM already produces an insecure completion for $X_i$, for two reasons:
(i) filtering $\mathcal{E}$ to model-specific failures would make the editing set model-dependent and hinder fair cross-model comparison;
(ii) including all instances allows us to quantify editing side effects such as functional regressions on already-secure contexts.

\vspace{-0.3cm}

\begin{table*}[b]
\centering
\caption{CWE categories in the dataset~\cite{He2023} and seen/unseen separation statistics. Severity is indicated by superscripts:
\textsuperscript{C}=Critical, \textsuperscript{H}=High, \textsuperscript{M}=Medium.
The severity levels are based on MITRE CWE data~\cite{MITRECorporation2024}.}
\vspace{-0.35cm}
\label{tab:sven_cwe}
\scriptsize
\begin{tabular}{l | l | c | c}
\toprule
\textbf{Split} & \textbf{CWE Types} & \textbf{Lang.} & \textbf{Size} \\
\midrule
\makecell[l]{Seen\\($\mathcal{T}_{seen}$)} &
\makecell[l]{%
CWE-089\textsuperscript{C}, CWE-125\textsuperscript{M}, CWE-078\textsuperscript{C},
CWE-476\textsuperscript{M}, CWE-416\textsuperscript{C}, CWE-022\textsuperscript{C},\\
CWE-787\textsuperscript{C}, CWE-079\textsuperscript{C}, CWE-190\textsuperscript{H}
} &
\makecell[c]{Python\\\& C/C++} &
\makecell[c]{710 train\\pairs + 24 eval\\prompts} \\
\midrule
\makecell[l]{Unseen\\($\mathcal{T}_{unseen}$)} &
\makecell[l]{%
CWE-119\textsuperscript{C}, CWE-502\textsuperscript{H}, CWE-732\textsuperscript{H}, CWE-798\textsuperscript{C},
CWE-020\textsuperscript{C}, CWE-094\textsuperscript{C},\\
CWE-116\textsuperscript{C}, CWE-117\textsuperscript{H},
CWE-209\textsuperscript{M}, CWE-215\textsuperscript{M}, CWE-312\textsuperscript{M}, CWE-327\textsuperscript{M},\\
CWE-377\textsuperscript{M}, CWE-611\textsuperscript{C}, CWE-643\textsuperscript{H}, CWE-777\textsuperscript{M},
CWE-918\textsuperscript{C}
} &
\makecell[c]{Python\\\& C/C++} &
\makecell[c]{29 eval\\prompts} \\
\midrule
\multicolumn{4}{p{\dimexpr\linewidth-2\tabcolsep\relax}}{%
\vspace{-\baselineskip}{\textit{\textbf{Seen/Unseen Separation:}}\quad
\# CWEs: 9 vs.\ 17 (Jaccard\,=\,0.0)\quad
CodeQL rules: 17 vs.\ 24 (Jaccard\,=\,0.0)\newline
Avg.\ prompt embedding similarity: 0.209 (seen)\,/\,0.244 (unseen)\,/\,0.167 (cross-set)}
} \\
\bottomrule
\end{tabular}
\vspace{-0.12cm}
\end{table*}

\subsection{Compared Methods}
We evaluate three representative model editing methods from different editing families and one SOTA secure code generation method.
Each model editing method receives the same security editing dataset $\mathcal{E}$ and produces an edited model $f_e = \mathcal{A}(f, \mathcal{E})$ via localized parameter updates; they differ in \textit{where} and \textit{how} these updates are applied.
\textbf{(1) UltraEdit~\cite{Gu2025}} is a local-modification method that computes a closed-form parameter shift from hidden states and output gradients at editable layers, avoiding iterative training. A lifelong normalization mechanism stabilizes feature distributions under sequential edits.
\textbf{(2) DINM~\cite{Wang2024a}} is a local-modification method that identifies domain-informative neurons via gradient-based attribution and applies targeted parameter shifts to steer outputs, achieving high specificity with limited impact on non-target behavior.
\textbf{(3) $\text{DEFER}^*$} adapts DEFER~\cite{Wang2024b, Hartvigsen2023}, an external-memorization method that stores edited knowledge in an adapter~\cite{Houlsby2019} with a router selecting between edited and base model predictions. 
Since sequence-level routing is less suitable for code generation where security cues appear at arbitrary token positions, we adopt the token-wise method that computes a routing score per token hidden state.
\textbf{(4) CoSec~\cite{Li2024, Li2025}} is an inference-time hardening method that pairs a target LLM with a lightweight auxiliary model trained to distinguish secure from insecure next-token predictions, applying a gating rule to bias decoding toward safer continuations without modifying model parameters.

\vspace{-0.2cm}
\subsection{Target LLMs}
We select six publicly available LLMs varying in scale, pretraining objectives, and model families: two general-purpose models, \textbf{LLaMA-3.1-8B} and \textbf{LLaMA-3.2-3B}~\cite{Grattafiori2024}, and four code-specialized models, \textbf{DeepSeek-Coder-6.7B}~\cite{Guo2024}, \textbf{CodeGen-6B}~\cite{Nijkamp2022}, \textbf{Qwen2.5-Coder-7B/3B}~\cite{Hui2024}, covering parameter scales from 3B to 8B across distinct model families.

\vspace{-0.25cm}
\subsection{Dataset}

\subsubsection{Hardening Dataset}

Following prior work~\cite{Li2024, Li2025, He2023}, we adopt the security-relevant dataset of He and Vechev~\cite{He2023}, comprising 710 vulnerable/fixed function pairs across 9 CWE categories in Python and C/C++ (see Table~\ref{tab:sven_cwe}).
For \textbf{CoSec}, we follow the official implementation~\cite{CoSecRepo} directly and use the original pairs of vulnerable and fixed functions to train the auxiliary security model.
For \textbf{model editing}, each pair is converted into an editing instance $M_i = (X_i, \hat{Y}_i, Y_i, \text{CWE}_i)$ through three steps:
(1) compute the syntactic diff using \texttt{difflib}~\cite{python-difflib} to locate the modified region;
(2) extract a vulnerability-relevant prefix $X_i$ (up to 150 tokens before the diff start) and a secure suffix $Y_i$ ($L$ tokens from the fixed function at the diff start);
(3) generate the original model's completion $\hat{Y}_i$ by prompting the target LLM with $X_i$.
The full dataset yields $|\mathcal{E}| = 710$ editing instances, covering all 9 CWE categories in $\mathcal{T}_{seen}$.
Each editing method consumes $\mathcal{E}$ in a single pass: instances are processed sequentially, and the cumulative effect of all edits produces the final edited model $f_e$.

\subsubsection{Evaluation Dataset}
Our study follows the three evaluation objectives defined in Section~\ref{sec:task_def}: Security Effectiveness, Security Generalization, and Functional Correctness.
We detail the selected datasets for each dimension.


\textbf{$\bullet$ Security Effectiveness.}
We adopt $D_{sec}$~\cite{Li2024, He2023}, a CWE-based evaluation suite covering nine categories in $\mathcal{T}_{seen}$, each instantiated into three fully specified prompts (including package declarations, variable definitions, and function signatures) designed to compile successfully.
For each scenario, the security-hardened model freely generates code completions. 
Valid completions (duplicates and non-compiling outputs removed) are analyzed by CodeQL~\cite{github_codeql} to detect vulnerabilities.

To examine robustness to realistic prompt variations~\cite{Zhan2024, Chen2024, sun2026cost}, we construct a perturbation set $D_{per}$ based on the \texttt{CWE-089 0-Py} scenario used in CoSec~\cite{Li2024}.
The perturbations cover metadata changes, documentation rewrites, and context-code edits, such as modifying import metadata, rewriting or removing comments, moving hard-coded parameters to environment variables, and switching database APIs.
These perturbations preserve the intended functionality while changing surface forms and surrounding context.
For each perturbed prompt, the model generates completions that are validated using CodeQL.
The full perturbation list is included in our replication package.


\textbf{$\bullet$ Security Generalization}.
To assess security generalization, we further evaluate the security-hardened models on unseen CWE categories.
Specifically, we use the 17 CWE types that do not appear in the training set (listed in Table~\ref{tab:sven_cwe}), denoted as $\mathcal{T}_{unseen}$.
This evaluation comprises 29 testing scenarios in total.
We verify that $\mathcal{T}_{seen}$ and $\mathcal{T}_{unseen}$ are distinct: their CWE sets and corresponding CodeQL rules are disjoint (Jaccard similarity = 0.0).
CodeBERT-based prompt embeddings further show higher intra-set similarity (0.209/0.244) than cross-set similarity (0.167), suggesting semantic separation between the two splits.

\textbf{$\bullet$ Functional Correctness}.
To investigate that model edits do not impair the LLM’s coding ability, we evaluate on the HumanEval benchmark~\cite{Chen2021}.
HumanEval contains 164 Python programming tasks, each with a function signature, natural-language description, and unit tests.
For each task, the model generates multiple completions that are then executed against the tests.

\vspace{-0.2cm}
\subsection{Evaluation Metrics}

We adopt two standard evaluation metrics widely used in secure code generation research~\cite{He2023, Li2023, Li2025} to assess both the security and functional correctness of model outputs.

\textbf{$\bullet$ Security Ratio (SR)}.
Let $N_t$, $N_d$, and $N_{np}$ denote the numbers of total completions, duplicates, and non-compilable samples, respectively; among the valid completions, $N_{sec}$ denotes those classified as secure by CodeQL.
The security ratio is defined as: 
\begin{equation}
\label{eq:sr}
\scriptsize
    SR = \frac{N_{sec}}{N_t - N_{np} - N_d}
\end{equation}

\textbf{$\bullet$ Pass@k}.
To measure functional correctness, we use the Pass@k metric following prior work~\cite{Chen2021}.
Given $n$ completions sampled for each task (with $n > k$), Pass@k estimates the probability that at least one of the model’s $k$ selected samples passes all reference tests:
\begin{equation}
\label{eq:passk}
\scriptsize
\text{Pass@}k := \mathbb{E}_{\text{Problems}}
\left[
1 -
\frac{\binom{n - c}{k}}{\binom{n}{k}}
\right]
\end{equation}
where $c$ denotes the number of completions (out of $n$) that successfully pass the unit tests.

\subsection{Evaluation Pipeline}

Our study aims to investigate whether model editing can serve as an effective model-level hardening mechanism for secure code generation.
To this end, we design a unified evaluation pipeline that assesses each hardening method along security effectiveness, generalization, functional correctness:

\textbf{\ding{172} Code Generation.}
For each prompt, the target LLM or its hardened variant generates 25 completions with nucleus sampling (Top-P\,=\,0.95), a maximum of 256 new tokens, and temperatures $T \in \{0.1, 0.4, 0.8\}$.
All experiments are repeated across 10 random seeds.
\textbf{\ding{173} Output Filtering.}
We remove exact duplicates and syntactically invalid outputs that fail to compile; only valid completions are evaluated.
\textbf{\ding{174} Security Evaluation.}
Valid completions are analyzed by CodeQL with CWE-specific queries.
A completion is classified as \emph{secure} if no query is triggered, and as \emph{vulnerable} otherwise.
SR is computed over valid completions following Equation~\ref{eq:sr}.
\textbf{\ding{175} Functional Correctness Evaluation.}
For HumanEval, completions are executed against the reference unit tests, and functional correctness is measured by Pass@$k$ ($k \in \{1, 5, 10\}$) following Equation~\ref{eq:passk}.

\subsection{Implementation Details}

All target LLMs and tokenizers are loaded from the official HuggingFace repositories.
We configure each hardening method as follows.
\textbf{$\bullet$ CoSec.}
Following Li \etal~\cite{Li2024}, we select the smallest model within each LLM family as the auxiliary security model: LLaMA-3.2-1B (for LLaMA-3.2-3B), LLaMA-3.1-8B (for LLaMA-3.1-8B), DeepSeek-Coder-1.3B (for DeepSeek-Coder-6.7B), CodeGen-350M (for CodeGen-6B), and Qwen2.5-Coder-0.5B (for Qwen-family models).
The auxiliary model is trained with LoRA (rank 8, $\alpha=16$, max length 1024, lr $5{\times}10^{-5}$, up to 8 epochs), with the best validation checkpoint selected.
\textbf{$\bullet$ UltraEdit.}
Following~\cite{Gu2025}, we apply UltraEdit to the \emph{second-to-last} Transformer layer via a single forward--backward pass.
\textbf{$\bullet$ DINM.}
We attach DINM to the \emph{fifth-to-last} Transformer layer with lr $5{\times}10^{-4}$, zero weight decay, and KL regularization factor $0.05$.
\textbf{$\bullet$ DEFER$^{*}$.}
We adapt DEFER to a token-wise variant inserted at the \emph{second-to-last} Transformer layer, optimized for 10 iterations (lr $7{\times}10^{-5}$).
At inference time, tokens with routing score above $0.5$ use the edited prediction; otherwise the original prediction is used.

For all editing methods, we set the target suffix length to $L=50$ unless otherwise specified.
All experiments are conducted on a single NVIDIA Tesla A800 80GB GPU with Intel 2.8GHz CPUs.

\vspace{-0.3cm}
\subsection{Research Questions}

In this work, we attempt to address the following Research Questions (RQs):
\vspace{-0.1cm}
\begin{description}
  \item[\textbf{RQ1: Security Effectiveness.}]
  To what extent do edits improve the security of generated code?
  \begin{description}
      \item[\textbf{RQ1.1}] To what extent do edited models improve security on seen CWEs?
      \item[\textbf{RQ1.2}] Are the security improvements robust to prompt perturbations?
  \end{description}
  \item[\textbf{RQ2: Security Generalization.}] 
  Do model edits improve security for unseen vulnerability types?
  \item[\textbf{RQ3: Functional Correctness.}] 
  Do the edits preserve the target model’s functional correctness?
  \item[\textbf{RQ4: Enhanced Approach.}]
  How can we further improve the model editing techniques for better performances on secure code generation?
\end{description}

\vspace{-0.3cm}
\section{Results and Analysis}
\subsection{RQ1: Security Effectiveness}
\subsubsection{Answer to RQ1.1}

\begin{table*}[t]
\centering
\vspace{-0.1cm}
\caption{Overall Security Ratio (\%) evaluation. The left section reports results on $D_{sec}$ (RQ1.1), and the right section reports results under prompt perturbations (RQ1.2).}
\label{tab:rq1_merged}
\vspace{-0.4cm}
\tiny
\setlength{\tabcolsep}{2.2pt} 
\begin{tabular}{ll cccc | cccc} 
\toprule
\multirow{2}{*}{\textbf{Model}} & \multirow{2}{*}{\textbf{Method}} & \multicolumn{4}{c}{\textbf{$D_{sec}$ (RQ1.1)}} & \multicolumn{4}{c}{\textbf{Perturbations (RQ1.2)}} \\
\cmidrule(lr){3-6} \cmidrule(lr){7-10}
 & & $T=0.1$ & $T=0.4$ & $T=0.8$ & Avg. & $T=0.1$ & $T=0.4$ & $T=0.8$ & Avg. \\
\midrule

\multirow{5}{*}{LLaMA3.1-8B}
 & Vanilla   & 68.6 & 68.8 & 65.8 & 67.7 & \underline{94.1} & \underline{91.1} & 76.6 & 87.3 \\
 & CoSec     & \underline{73.0 (+6.41\%)} & 77.4 (+12.50\%) & \underline{73.9 (+12.31\%)} & \underline{74.8} & 77.2 (-17.96\%) & 80.4 (-11.75\%) & 75.6 (-1.31\%) & 77.7 \\
 & DEFER* & 71.8 (+4.66\%) & 69.5 (+1.02\%) & 63.9 (-2.89\%) & 68.4 & 94.1 (+0.00\%) & 84.4 (-7.35\%) & 71.6 (-6.53\%) & 83.4 \\
 & DINM      & \cbg\textbf{81.2 (+18.37\%)} & \cbg\textbf{81.2 (+18.02\%)} & \cbg\textbf{75.2 (+14.29\%)} & \cbg\textbf{79.2} & 85.5 (-9.14\%) & \cbg\textbf{94.0 (+3.18\%)} & \cbg\textbf{94.9 (+23.89\%)} & \cbg\textbf{91.5} \\
 & UltraEdit & 71.9 (+4.81\%) & \underline{78.3 (+13.81\%)} & 70.4 (+6.99\%) & 73.5 & \cbg\textbf{97.6 (+3.72\%)} & 88.3 (-3.07\%) & \underline{79.4 (+3.66\%)} & \underline{88.4} \\
\midrule

\multirow{5}{*}{LLaMA3.2-3B}
 & Vanilla   & 61.3 & 62.2 & 64.8 & 62.8 & \cbg\textbf{100.0} & 93.5 & 81.5 & 91.7 \\
 & CoSec     & 65.7 (+7.18\%) & 67.4 (+8.36\%) & 63.6 (-1.85\%) & 65.6 & 75.7 (-24.30\%) & 81.4 (-12.94\%) & 77.9 (-4.42\%) & 78.3 \\
 & DEFER* & 61.3 (+0.00\%) & 64.3 (+3.38\%) & 63.5 (-2.01\%) & 63.0 & \cbg\textbf{100.0 (+0.00\%)} & 93.3 (-0.21\%) & 77.2 (-5.28\%) & 90.2 \\
 & DINM      & \underline{70.6 (+15.17\%)} & \cbg\textbf{76.2 (+22.51\%)} & \cbg\textbf{77.2 (+19.14\%)} & \cbg\textbf{74.7} & \cbg\textbf{100.0 (+0.00\%)} & \cbg\textbf{100.0 (+6.95\%)} & \cbg\textbf{97.8 (+20.00\%)} & \cbg\textbf{99.3} \\
 & UltraEdit & \cbg\textbf{70.8 (+15.50\%)} & \underline{70.0 (+12.54\%)} & \underline{65.9 (+1.70\%)} & \underline{68.9} & \cbg\textbf{100.0 (+0.00\%)} & \underline{95.1 (+1.71\%)} & \underline{88.4 (+8.47\%)} & \underline{94.5} \\
\midrule

\multirow{5}{*}{DeepSeek-6.7B}
 & Vanilla   & 77.8 & 71.3 & 70.6 & 73.2 & \cbg\textbf{94.1} & 92.9 & 84.1 & 90.4 \\
 & CoSec     & 77.8 (+0.00\%) & 73.5 (+3.09\%) & 74.5 (+5.52\%) & 75.3 & 93.3 (-0.85\%) & \underline{94.4 (+1.61\%)} & 92.8 (+10.34\%) & 93.5 \\
 & DEFER* & \cbg\textbf{83.7 (+7.58\%)} & \cbg\textbf{83.5 (+17.11\%)} & \cbg\textbf{82.2 (+16.43\%)} & \cbg\textbf{83.1} & \cbg\textbf{94.1 (+0.00\%)} & 94.1 (+1.29\%) & \cbg\textbf{94.6 (+12.49\%)} & \underline{94.3} \\
 & DINM      & \underline{82.0 (+5.40\%)} & \underline{81.2 (+13.88\%)} & \underline{77.4 (+9.63\%)} & \underline{80.2} & \cbg\textbf{94.1 (+0.00\%)} & \cbg\textbf{95.2 (+2.48\%)} & \underline{94.4 (+12.25\%)} & \cbg\textbf{94.6} \\
 & UltraEdit & 81.0 (+4.11\%) & 76.7 (+7.57\%) & 70.2 (-0.57\%) & 76.0 & \cbg\textbf{94.1 (+0.00\%)} & 91.4 (-1.61\%) & 76.5 (-9.04\%) & 87.3 \\
\midrule

\multirow{5}{*}{Qwen2.5-C-7B}
 & Vanilla   & 76.4 & 79.5 & 80.5 & 78.8 & \cbg\textbf{94.1} & 94.1 & 92.0 & 93.4 \\
 & CoSec     & 75.7 (-0.92\%) & 76.6 (-3.65\%) & 76.7 (-4.72\%) & 76.3 & 88.9 (-5.53\%) & 87.7 (-6.80\%) & 88.0 (-4.35\%) & 88.2 \\
 & DEFER* & 68.9 (-9.82\%) & \underline{83.9 (+5.53\%)} & \cbg\textbf{90.7 (+12.67\%)} & 81.2 & 88.2 (-6.27\%) & 94.1 (+0.00\%) & 91.6 (-0.43\%) & 91.3 \\
 & DINM      & \cbg\textbf{83.2 (+8.90\%)} & \cbg\textbf{83.9 (+5.53\%)} & 80.5 (+0.00\%) & \cbg\textbf{82.5} & \underline{91.2 (-3.08\%)} & \cbg\textbf{96.3 (+2.34\%)} & \cbg\textbf{96.2 (+4.57\%)} & \cbg\textbf{94.6} \\
 & UltraEdit & \underline{82.9 (+8.51\%)} & 80.3 (+1.01\%) & \underline{82.4 (+2.36\%)} & \underline{81.9} & \cbg\textbf{94.1 (+0.00\%)} & \underline{94.1 (+0.00\%)} & \underline{93.1 (+1.20\%)} & \underline{93.8} \\
\midrule

\multirow{5}{*}{CodeGen-6B}
 & Vanilla   & 67.0 & 65.4 & 68.0 & 66.8 & 93.3 & 78.5 & 63.2 & 78.3 \\
 & CoSec     & 71.3 (+6.42\%) & 69.8 (+6.73\%) & 71.6 (+5.29\%) & 70.9 & 76.5 (-18.01\%) & 80.8 (+1.91\%) & 83.6 (+32.28\%) & 80.0 \\
 & DEFER* & 22.2 (-66.87\%) & 53.7 (-17.89\%) & \cbg\textbf{82.8 (+21.76\%)} & 52.9 & 41.2 (-55.84\%) & 67.7 (-13.76\%) & 81.2 (+28.48\%) & 63.4 \\
 & DINM      & \cbg\textbf{80.6 (+20.30\%)} & \cbg\textbf{84.0 (+28.44\%)} & \underline{76.7 (+12.79\%)} & \cbg\textbf{80.4} & \underline{98.0 (+5.04\%)} & \underline{96.2 (+22.55\%)} & \underline{90.6 (+43.35\%)} & \underline{94.9} \\
 & UltraEdit & \underline{75.8 (+13.13\%)} & \underline{71.0 (+8.56\%)} & 76.5 (+12.50\%) & \underline{74.4} & \cbg\textbf{100.0 (+7.18\%)} & \cbg\textbf{98.3 (+25.22\%)} & \cbg\textbf{99.1 (+56.80\%)} & \cbg\textbf{99.1} \\
\midrule

\multirow{5}{*}{Qwen2.5-C-3B}
 & Vanilla   & \cbg\textbf{80.3} & \underline{75.0} & 67.1 & \underline{74.1} & 94.1 & 93.7 & 83.8 & 90.5 \\
 & CoSec     & 64.9 (-19.18\%) & 62.0 (-17.33\%) & 65.8 (-1.94\%) & 64.2 & 68.8 (-26.89\%) & 68.5 (-26.89\%) & 70.5 (-15.87\%) & 69.3 \\
 & DEFER* & 79.0 (-1.62\%) & 71.5 (-4.67\%) & 66.9 (-0.30\%) & 72.5 & 88.2 (-6.27\%) & 92.3 (-1.49\%) & 87.2 (+4.06\%) & 89.2 \\
 & DINM      & 71.9 (-10.46\%) & 70.2 (-6.40\%) & \cbg\textbf{72.1 (+7.45\%)} & 71.4 & \cbg\textbf{96.1 (+2.13\%)} & \cbg\textbf{96.2 (+2.67\%)} & \cbg\textbf{97.2 (+15.99\%)} & \cbg\textbf{96.5} \\
 & UltraEdit & \underline{79.4 (-1.12\%)} & \cbg\textbf{77.9 (+3.87\%)} & \underline{68.8 (+2.53\%)} & \cbg\textbf{75.4} & \underline{94.1 (+0.00\%)} & \underline{94.1 (+0.43\%)} & \underline{91.6 (+9.31\%)} & \underline{93.3} \\
\bottomrule
\end{tabular}
\vspace{-0.2cm}
\end{table*}



Table~\ref{tab:rq1_merged} reports the SRs of all methods on the seen CWE scenarios ($D_{sec}$) under three decoding temperatures (0.1, 0.4, and 0.8), where \textit{vanilla} denotes the target model without any security hardening.
Three key observations can be drawn from the results.

(1) \textit{Model editing methods generally provide larger and more stable security gains than CoSec}.
Across all models and temperatures, DINM delivers the strongest overall improvement (+11.27\% on average), followed by UltraEdit (+6.54\%), whereas CoSec improves security by only +1.46\% on average.
For instance, DINM raises LLaMA3.1-8B from 67.7\% to 79.2\% and CodeGen-6B from 66.8\% to 80.4\%, with improvements observed consistently across temperatures.
UltraEdit also outperforms vanilla models without harming performance.
Overall, these results suggest that injecting security-relevant knowledge into internal representations is more effective than inference-time guidance for improving code security.


(2) \textit{CoSec’s improvements are limited and even negative}.
CoSec delivers less consistent security improvements than model editing, and in several settings even underperforms the vanilla model.
We attribute this limitation to three factors.
\ding{182}
CoSec often relies on a much smaller auxiliary model (\eg CodeGen-350M for CodeGen-6B and Qwen2.5-Coder-0.5B for Qwen models), which limits its ability to capture fine-grained vulnerability cues in complex contexts; this is particularly evident on Qwen-family models, where CoSec consistently lags behind vanilla across temperatures.
CoSec produces meaningful improvements mainly when the auxiliary model is close to the target in architecture/scale (\eg LLaMA3.1-8B$\rightarrow$LLaMA3.1-8B), but this also increases latency due to dual-model decoding.
\ding{183}
CoSec does not modify the security-relevant knowledge of the target model; it only reweights token probabilities at inference time, so unsafe generation tendencies can persist.
\ding{184}
Token-level biasing may suppress obviously insecure tokens, but can also penalize benign or security-neutral continuations.

(3) \textit{Security improvements are not always guaranteed, especially at low decoding temperatures}.
At low temperatures (\eg $T=0.1$), both model editing and CoSec can occasionally reduce security, especially on Qwen2.5-Coder-3B.
Low-temperature decoding may resurface vulnerable patterns when security-relevant representations are only partially corrected, suggesting that smaller models may lack sufficient capacity to reliably absorb injected security knowledge.


\myfinding{1.1}{
Across all target LLMs, model editing consistently outperforms CoSec in security improvement.
DINM achieves the largest average gain (+11.27\%), followed by UltraEdit (+6.54\%), while CoSec yields smaller and sometimes negative gains, suggesting that parameter-level updates are more effective than inference-time steering.
}

\vspace{-0.2cm}
\subsubsection{Answer to RQ1.2}

Table~\ref{tab:rq1_merged} reports the SRs of all methods under a suite of prompt perturbations.
These perturbations emulate realistic variations in developer-written prompts that often destabilize the security behavior of LLMs.
The results reveal several noteworthy observations.

(1) \textit{Model editing remains more effective and stable than CoSec under perturbations}. UltraEdit and DINM maintain high SR across perturbed prompts, indicating that their security gains are not tied to specific prompt patterns.
(2) \textit{CoSec is brittle}. It often degrades SR across LLMs, because its parameter-free co-decoding guidance can be weakened by small context changes, causing models to revert to insecure code.
(3) \textit{Editing robustness varies by method and model}. DINM and UltraEdit are consistently strong on CodeGen-6B and LLaMA3.2-3B, whereas DEFER* is less stable.
(4) \textit{Perturbation robustness aligns with RQ1.1}. Models that gain more security from editing in the original CWE scenarios also tend to remain robust under prompt perturbations.

\myfinding{1.2}{
Most model editing methods preserve the effectiveness of security hardening under prompt perturbations.
In contrast, CoSec exhibits inconsistent behavior and often performs worse than the vanilla model when small perturbations are applied.
}
\vspace{-0.3cm}
\subsection{RQ2: Security Generalization}

\begin{table*}[h]
\centering
\caption{Security ratio (\%) on unseen CWE types (RQ2). Each cell reports absolute performance and relative change over the vanilla model at the same temperature.}
\label{tab:rq2_unseen}
\scriptsize
\vspace{-0.30cm}

\setlength{\tabcolsep}{3.5pt}
\renewcommand{\arraystretch}{1.05}

\resizebox{\textwidth}{!}{%
\begin{tabular}{l l c c c c @{\hspace{8pt}} l l c c c c}
\toprule
\textbf{Model} & \textbf{Method} & \textbf{$T=0.1$} & \textbf{$T=0.4$} & \textbf{$T=0.8$} & \textbf{Avg.} &
\textbf{Model} & \textbf{Method} & \textbf{$T=0.1$} & \textbf{$T=0.4$} & \textbf{$T=0.8$} & \textbf{Avg.} \\
\midrule

\multirow{5}{*}{\makecell[l]{LLaMA\\3.1-8B}}
 & Vanilla   & 51.0 & 54.9 & 54.9 & 53.6
& \multirow{5}{*}{\makecell[l]{LLaMA\\3.2-3B}}
 & Vanilla   & 45.5 & 49.4 & 53.1 & 49.3 \\
 & CoSec     & \underline{58.1 (+13.92\%)} & \cbg\textbf{60.4 (+10.02\%)} & 59.3 (+8.01\%) & 59.3
&  & CoSec     & \cbg\textbf{53.8 (+18.24\%)} & \underline{55.1 (+11.54\%)} & \underline{55.7 (+4.90\%)} & \underline{54.9} \\
 & DEFER*    & 57.0 (+11.76\%) & 58.9 (+7.29\%) & \cbg\textbf{62.6 (+14.03\%)} & \underline{59.5}
&  & DEFER*    & \underline{50.3 (+10.55\%)} & 50.1 (+1.42\%) & 55.0 (+3.58\%) & 51.8 \\
 & DINM      & \cbg\textbf{61.3 (+20.20\%)} & \underline{59.7 (+8.74\%)} & \underline{60.8 (+10.75\%)} & \cbg\textbf{60.6}
&  & DINM      & 48.9 (+7.47\%) & \cbg\textbf{57.2 (+15.79\%)} & \cbg\textbf{64.0 (+20.53\%)} & \cbg\textbf{56.7} \\
 & UltraEdit & 51.1 (+0.20\%) & 54.9 (+0.00\%) & 55.2 (+0.55\%) & 53.7
&  & UltraEdit & 44.9 (-1.32\%) & 49.9 (+1.01\%) & 51.7 (-2.64\%) & 48.8 \\
\midrule

\multirow{5}{*}{\makecell[l]{DeepSeek\\-6.7B}}
 & Vanilla   & 55.8 & 58.1 & 58.3 & 57.4
& \multirow{5}{*}{\makecell[l]{Qwen2.5\\-Coder-7B}}
 & Vanilla   & \cbg\textbf{57.4} & 59.4 & 59.1 & \underline{58.6} \\
 & CoSec     & \underline{62.2 (+11.47\%)} & 58.0 (-0.17\%) & 56.1 (-3.77\%) & \underline{58.8}
&  & CoSec     & 56.5 (-1.57\%) & 56.1 (-5.56\%) & 57.0 (-3.55\%) & 56.5 \\
 & DEFER*    & 53.7 (-3.76\%) & 56.9 (-2.07\%) & \underline{59.1 (+1.37\%)} & 56.6
&  & DEFER*    & 33.8 (-41.11\%) & \cbg\textbf{62.4 (+5.05\%)} & \underline{59.6 (+0.85\%)} & 51.9 \\
 & DINM      & 57.9 (+3.76\%) & \underline{59.0 (+1.55\%)} & 57.7 (-1.03\%) & 58.2
&  & DINM      & 56.6 (-1.39\%) & \underline{60.8 (+2.36\%)} & \cbg\textbf{61.4 (+3.89\%)} & \cbg\textbf{59.6} \\
 & UltraEdit & \cbg\textbf{65.3 (+17.03\%)} & \cbg\textbf{64.1 (+10.33\%)} & \cbg\textbf{65.6 (+12.52\%)} & \cbg\textbf{65.0}
&  & UltraEdit & \underline{57.0 (-0.70\%)} & 58.5 (-1.52\%) & 58.5 (-1.02\%) & 58.0 \\
\midrule

\multirow{5}{*}{\makecell[l]{CodeGen\\-6B}}
 & Vanilla   & 43.4 & 51.0 & 54.9 & 49.8
& \multirow{5}{*}{\makecell[l]{Qwen2.5\\-Coder-3B}}
 & Vanilla   & \underline{64.2} & 62.2 & 57.2 & 61.2 \\
 & CoSec     & \underline{51.5 (+18.66\%)} & \underline{54.3 (+6.47\%)} & 54.5 (-0.73\%) & \underline{53.4}
&  & CoSec     & 63.2 (-1.56\%) & \underline{62.6 (+0.64\%)} & \underline{60.5 (+5.77\%)} & \underline{62.1} \\
 & DEFER*    & 28.7 (-33.87\%) & 46.1 (-9.61\%) & \underline{68.1 (+24.04\%)} & 47.6
&  & DEFER*    & 48.3 (-24.77\%) & 52.7 (-15.27\%) & 54.1 (-5.42\%) & 51.7 \\
 & DINM      & 42.9 (-1.15\%) & 46.9 (-8.04\%) & 60.2 (+9.65\%) & 50.0
&  & DINM      & \cbg\textbf{64.4 (+0.31\%)} & \cbg\textbf{66.2 (+6.43\%)} & \cbg\textbf{66.6 (+16.43\%)} & \cbg\textbf{65.7} \\
 & UltraEdit & \cbg\textbf{53.0 (+22.12\%)} & \cbg\textbf{64.1 (+25.69\%)} & \cbg\textbf{78.9 (+43.72\%)} & \cbg\textbf{65.3}
&  & UltraEdit & 62.7 (-2.34\%) & 62.3 (+0.16\%) & 55.4 (-3.15\%) & 60.1 \\
\bottomrule
\end{tabular}%
}
\vspace{-0.10cm}
\end{table*}

Table~\ref{tab:rq2_unseen} reports the security ratios on 17 unseen CWE categories under three temperatures, evaluating whether security improvements can transfer beyond the vulnerability types used for training or editing. Overall, most methods, including CoSec, improve over vanilla models, suggesting that vulnerability knowledge can generalize to some extent. Interestingly, unlike its limited and unstable gains on seen CWEs in RQ1, CoSec improves on unseen CWEs, indicating that its auxiliary model may capture vulnerability-agnostic signals rather than specific CWE patterns.

Model editing methods exhibit more varied transferability. DINM remains strong across most LLMs, whereas DEFER* achieves lower SR, showing that not all editing mechanisms produce transferable security knowledge. This suggests that generalization depends on both the editing mechanism and the editing data. For example, UltraEdit’s sparse up-projection edits can effectively suppress seen vulnerable patterns but may bind the injected behavior to edited contexts, limiting transfer to structurally different CWEs. In contrast, DINM’s gradient-attribution-based neuron updates may capture broader security-related semantics, leading to more consistent transfer. Nevertheless, the improvements over vanilla models indicate that model editing is not merely memorization and can transfer part of the injected security knowledge to unseen CWE scenarios.


\myfinding{2}{
RQ2 shows that strong security improvements on seen vulnerabilities do not necessarily transfer to unseen ones.
Model editing often yields larger improvements on seen CWEs, but these gains may diminish on unseen categories.
On the other hand, CoSec improves less on seen CWEs yet often stays close to or slightly above the vanilla model on unseen cases.
}

\subsection{RQ3: Functional Correctness}
\label{sec:rq3}

\begin{table*}[h]
\centering
\vspace{-0.1cm}
\caption{Functional correctness measured by Pass@1 (P@1), Pass@5 (P@5), and Pass@10 (P@10) under different decoding temperatures (RQ3). Best results are highlighted in bold with gray background, and second-best results are underlined (ties are marked accordingly). Method abbreviations: Van.=Vanilla, CoS.=CoSec, DEF.=DEFER*, UE=UltraEdit.}
\label{tab:rq3_pass}
\vspace{-0.28cm}
\tiny
\setlength{\tabcolsep}{1.5pt}
\renewcommand{\arraystretch}{1.2}
\begin{tabular}{llccccccccc||llccccccccc}
\toprule
\multirow{2}{*}{\textbf{Model}} & \multirow{2}{*}{\textbf{Met.}} &
\multicolumn{3}{c}{$T{=}0.1$} &
\multicolumn{3}{c}{$T{=}0.4$} &
\multicolumn{3}{c||}{$T{=}0.8$} &
\multirow{2}{*}{\textbf{Model}} & \multirow{2}{*}{\textbf{Met.}} &
\multicolumn{3}{c}{$T{=}0.1$} &
\multicolumn{3}{c}{$T{=}0.4$} &
\multicolumn{3}{c}{$T{=}0.8$} \\
\cmidrule(lr){3-5}\cmidrule(lr){6-8}\cmidrule(lr){9-11}
\cmidrule(lr){14-16}\cmidrule(lr){17-19}\cmidrule(lr){20-22}
& & P@1 & P@5 & P@10 & P@1 & P@5 & P@10 & P@1 & P@5 & P@10 &
& & P@1 & P@5 & P@10 & P@1 & P@5 & P@10 & P@1 & P@5 & P@10 \\
\midrule

\multirow{5}{*}{LLaMA3.1-8B}
& Van.  & \second{38.1} & \second{47.9} & \second{51.1} & \best{36.7}   & \second{57.9} & \second{66.2} & \second{31.9} & \second{57.3} & \second{67.0} &
\multirow{5}{*}{Qwen2.5-C-7B}
& Van.  & \second{82.7} & 86.3 & 87.0 & \best{82.2} & \second{90.9} & \second{92.7} & \best{79.3} & \second{93.7} & \second{95.9} \\
& CoS.  & 35.4 & \best{57.8} & \best{66.0} & \second{36.1} & \best{59.4} & \best{68.0} & \best{33.8} & \best{60.3} & \best{70.6} &
& CoS.  & 75.3 & \best{94.6} & \best{96.7} & 75.2 & \best{94.7} & \best{96.7} & 75.1 & \best{94.6} & \best{96.6} \\
& DEF.  & 34.9 & 42.2 & 44.1 & 31.4 & 48.4 & 53.5 & 26.3 & 50.0 & 59.7 &
& DEF.  & 74.5 & 79.0 & 79.8 & 72.2 & 86.8 & 89.5 & 65.7 & 89.0 & 92.6 \\
& DINM  & 24.0 & 28.0 & 29.0 & 20.1 & 32.2 & 37.2 & 12.6 & 28.4 & 36.1 &
& DINM  & 76.2 & 80.7 & 82.0 & 75.5 & 89.1 & 91.7 & 64.6 & 90.4 & 94.3 \\
& UE    & \best{38.9} & 46.4 & 48.7 & 35.7 & 54.3 & 61.1 & 30.3 & 55.5 & 65.5 &
& UE    & \best{83.0} & \second{86.6} & \second{87.3} & \second{81.9} & \second{90.9} & \second{92.7} & \second{79.1} & 93.4 & 95.7 \\
\midrule

\multirow{5}{*}{LLaMA3.2-3B}
& Van.  & \second{26.7} & \second{31.5} & 33.1 & \second{26.7} & \best{40.1} & \best{46.5} & \best{24.0} & \best{43.1} & \best{52.4} &
\multirow{5}{*}{CodeGen-6B}
& Van.  & \best{18.6} & \second{20.7} & \second{21.2} & \best{18.1} & \best{24.8} & \best{27.6} & \best{14.0} & \best{25.1} & \best{29.7} \\
& CoS.  & 24.1 & \best{40.2} & \best{47.7} & 22.8 & 38.4 & 45.6 & 19.8 & 36.8 & 45.1 &
& CoS.  & 12.4 & \best{22.5} & \best{27.0} & 12.3 & \second{22.2} & \second{26.4} & \second{10.6} & \second{20.3} & \second{24.7} \\
& DEF.  & 26.4 & 31.3 & 33.1 & \best{26.9} & \second{40.0} & 45.0 & 20.8 & 39.4 & 48.0 &
& DEF.  & 11.2 & 15.1 & 16.1 & 10.6 & 17.4 & 19.7 & 9.3  & 18.2 & 21.7 \\
& DINM  & 12.8 & 15.5 & 16.7 & 10.2 & 18.2 & 21.5 & 6.3  & 14.9 & 19.3 &
& DINM  & \second{12.9} & 15.1 & 16.1 & \second{12.7} & 17.7 & 19.9 & 10.2 & 18.4 & 22.0 \\
& UE    & \best{27.1} & 31.4 & \second{33.3} & 26.5 & 39.7 & \second{46.0} & \second{22.9} & \second{42.0} & \second{51.2} &
& UE    & 10.4 & 14.5 & 16.4 & 8.6  & 17.4 & 21.1 & 2.5  & 7.9  & 11.2 \\
\midrule

\multirow{5}{*}{DeepSeek-6.7B}
& Van.  & \best{46.7} & \second{54.6} & \second{57.1} & \best{43.6} & \second{64.2} & \second{70.8} & \best{39.0} & \best{66.7} & \best{75.1} &
\multirow{5}{*}{Qwen2.5-C-3B}
& Van.  & \second{59.6} & \second{67.5} & \second{70.1} & \best{58.2} & \best{78.5} & \second{83.3} & \second{53.3} & \best{80.2} & \best{86.1} \\
& CoS.  & 38.8 & \best{65.2} & \best{73.7} & 39.2 & \best{64.9} & \best{73.2} & 35.2 & 63.5 & 73.0 &
& CoS.  & 43.1 & \best{77.0} & \best{86.4} & 42.7 & 76.3 & \best{85.7} & 42.5 & 76.0 & \second{85.6} \\
& DEF.  & 37.2 & 47.2 & 50.0 & 38.9 & 60.3 & 66.5 & \second{36.1} & \second{65.5} & \second{74.6} &
& DEF.  & 51.4 & 59.3 & 61.6 & 49.3 & 71.4 & 77.5 & 43.7 & 73.6 & 80.7 \\
& DINM  & \second{44.3} & 52.3 & 54.4 & \second{42.4} & 62.2 & 68.6 & 36.0 & 63.8 & 72.1 &
& DINM  & 51.5 & 60.7 & 63.0 & \second{51.0} & 72.3 & 77.8 & 43.4 & 73.9 & 81.8 \\
& UE    & 31.1 & 41.0 & 44.5 & 32.4 & 56.2 & 65.2 & 25.0 & 54.3 & 65.9 &
& UE    & \best{60.3} & 67.4 & 69.5 & \best{58.2} & \second{77.5} & 81.4 & \best{53.7} & \second{80.1} & 85.4 \\
\bottomrule
\end{tabular}
\vspace{-0.20cm}
\end{table*}


Table~\ref{tab:rq3_pass} reports functional correctness under different decoding temperatures.
Overall, both model editing and CoSec introduce non-trivial side effects, revealing an inherent tension between security hardening and functional correctness.
(1) \textit{Model editing generally induces larger correctness degradation than CoSec}.
DINM consistently reduces Pass@1/Pass@5 across most models and temperatures (\eg on LLaMA3.1-8B at $T=0.1$, Pass@1 drops from 38.1\% to 24.0\%), suggesting that parameter modification may disrupt shared representations needed for general code synthesis.
(2) \textit{UltraEdit exhibits a different correctness profile}.
Unlike DINM, UltraEdit can slightly improve Pass@1 at low temperatures (\eg LLaMA3.1-8B: 38.9\% vs.\ 38.1\%), yet these gains diminish at higher temperatures where it often trails CoSec on Pass@5/Pass@10.
(3) \textit{At higher temperatures, CoSec tends to preserve correctness better}.
When decoding is more stochastic ($T=0.4$/$0.8$), CoSec frequently achieves the best or second-best Pass@5/Pass@10, particularly on larger models.
This is consistent with inference-time steering: reshaping token probabilities without permanently altering internal representations keeps correct solutions accessible under diverse sampling.

\begin{table*}[t]
\caption{The trade-off score aggregated from security ratio and functional correctness across decoding temperatures.
\textbf{Bold} indicates the best score among the three editing methods for each model and temperature.}
\vspace{-0.3cm}
\label{tab:harmonic_tradeoff_horizontal}
\centering
\tiny
\setlength{\tabcolsep}{1.8pt}
\begin{tabular}{llccc c llccc c llccc}
\toprule
\textbf{Model} & \textbf{Method} & $T=0.1$ & $T=0.4$ & $T=0.8$ & &
\textbf{Model} & \textbf{Method} & $T=0.1$ & $T=0.4$ & $T=0.8$ & &
\textbf{Model} & \textbf{Method} & $T=0.1$ & $T=0.4$ & $T=0.8$ \\
\midrule
\multirow{5}{*}{\shortstack{LLaMA\\3.1-8B}}
 & Vanilla   & 48.99 & 47.87 & 42.97 & &
\multirow{5}{*}{\shortstack{LLaMA\\3.2-3B}}
 & Vanilla   & 37.20 & 37.36 & 35.03 & &
\multirow{5}{*}{\shortstack{DeepSeek\\6.7B}}
 & Vanilla   & 58.37 & 54.11 & 50.24 \\
 & CoSec     & 47.68 & 49.24 & 46.38 & &
 & CoSec     & 35.26 & 34.07 & 30.20 & &
 & CoSec     & 51.78 & 51.13 & 47.81 \\
 & DEFER*    & 46.97 & 43.26 & 37.26 & &
 & DEFER*    & 36.91 & 37.93 & 31.34 & &
 & DEFER*    & 51.51 & 53.07 & \textbf{50.17} \\
 & DINM      & 37.05 & 32.22 & 21.58 & &
 & DINM      & 21.67 & 17.99 & 11.65 & &
 & DINM      & \textbf{57.52} & \textbf{55.71} & 49.14 \\
 & UltraEdit & \textbf{50.49} & \textbf{49.04} & \textbf{42.37} & &
 & UltraEdit & \textbf{39.20} & \textbf{38.45} & \textbf{33.99} & &
 & UltraEdit & 44.94 & 45.56 & 36.87 \\
\midrule
\multirow{5}{*}{\shortstack{Qwen2.5\\Cod-7B}}
 & Vanilla   & 79.43 & 80.83 & 79.90 & &
\multirow{5}{*}{\shortstack{CodeGen\\6B}}
 & Vanilla   & 29.12 & 28.35 & 23.22 & &
\multirow{5}{*}{\shortstack{Qwen2.5\\Cod-3B}}
 & Vanilla   & 68.42 & 65.54 & 59.41 \\
 & CoSec     & 75.50 & 75.89 & 75.89 & &
 & CoSec     & 21.13 & 20.91 & 18.47 & &
 & CoSec     & 51.80 & 50.57 & 51.64 \\
 & DEFER*    & 71.59 & 77.61 & 76.20 & &
 & DEFER*    & 14.89 & 17.71 & 16.72 & &
 & DEFER*    & 62.28 & 58.36 & 52.87 \\
 & DINM      & 79.55 & 79.48 & 71.68 & &
 & DINM      & \textbf{22.24} & \textbf{22.06} & \textbf{18.01} & &
 & DINM      & 60.01 & 59.08 & 54.18 \\
 & UltraEdit & \textbf{82.95} & \textbf{81.09} & \textbf{80.72} & &
 & UltraEdit & 18.29 & 15.34 & 4.84 & &
 & UltraEdit & \textbf{68.54} & \textbf{66.62} & \textbf{60.32} \\
\bottomrule
\end{tabular}
\end{table*}

\textbf{Connecting RQ1 and RQ3.}
Since security and correctness can trade off against each other, reporting either metric in isolation risks misleading conclusions, as a method may appear superior by sacrificing the other dimension.
We therefore compute a harmonic-mean trade-off score $\text{TradeOff} = {2 \times \text{SR} \times \text{Pass@1}} / {(\text{SR} + \text{Pass@1})}$, analogous to F1-score, which jointly rewards both dimensions and penalizes methods excelling in only one.
Table~\ref{tab:harmonic_tradeoff_horizontal} shows that model editing methods often sacrifice correctness for security gains.
This effect is particularly evident for aggressive editing strategies such as DINM, whose parameter modifications substantially distort representations reused for general code synthesis, especially under cumulative edits.
CoSec, by contrast, better preserves correctness through inference-time steering at the cost of limited peak security improvements.
Among all editing approaches, UltraEdit often achieves the highest or near-highest trade-off scores across most models and temperatures, ranking best in 12 out of 18 configurations (66.7\%).

\myfinding{3}{
Overall, security hardening is not free: model editing typically yields larger security gains but risks greater correctness regressions, whereas CoSec offers more conservative gains with better functional preservation. Among editing methods, UltraEdit achieves the best balance, ranking first in 12/18 configurations.
}

\subsection{RQ4: Performance Improvement}

\textbf{Approach.}
UltraEdit achieves the most favorable trade-off in our evaluation and remains highly efficient (Section~\ref{sec:efficiency}).
Yet its edits still impair general programming capability, motivating us to refine UltraEdit to better preserve functional correctness without weakening its security benefits.
To this end, we propose \toolname{}, a post-edit refinement method for models edited by UltraEdit.
The key challenge is that standard fine-tuning for functional recovery can overwrite the security-relevant parameter changes introduced by editing, since both processes modify overlapping regions of the model.
We refer to this as the \emph{security edit preservation} problem: how to restore functional correctness without erasing injected security knowledge.
To address this, \toolname{} combines:
(1) \textbf{\textit{functional tuning}} on instruction-style code generation data, and
(2) \textbf{\textit{edit-aware regularization}} that constrains the edited parameters to stay close to their post-edit values.

Formally, let $f$ denote the target LLM and $f_e$ be the model produced by \textsc{UltraEdit}.
Let $\theta_{edit}$ denote the subset of parameters modified by UltraEdit and $\theta_{edit}^{e}$ their post-edit values; unlike standard regularization approaches that anchor task-performance-critical parameters~\cite{Kirkpatrick2017}, \toolname{} anchors precisely $\theta_{edit}$, ensuring that functional recovery does not overwrite the injected security behavior.
Given an instruction--code pair $(i,c)\in\mathcal{D}_{std}$, we perform standard next-token training:
\begin{equation}
\small
\mathcal{L}_{std}(i,c) = -\sum_{t=1}^{|c|} \log P(c_t \mid c_{<t}, i).
\end{equation}
To preserve the injected security behavior, we add an edit-aware regularization term:
\begin{equation}
\small
\mathcal{L}_{reg} = \lVert \theta_{edit} - \theta_{edit}^{e} \rVert_2^2,
\end{equation}
The final loss function is:
\begin{equation}
\small
\mathcal{L}=\mathcal{L}_{std}+ \mu \mathcal{L}_{reg}.
\end{equation}
where $\mu$ controls the regularization strength (default $\mu = 1\times10^{-3}$; sensitivity analysis in Section~\ref{sec:sensitivity}).
For functional tuning, we use \textit{Code Evol-Instruct}~\cite{Luo2023} as the supervision source, which provides diverse instruction-style code generation pairs covering general programming tasks without overlap with $\mathcal{E}$, thereby avoiding contamination of the security evaluation.
We fine-tune $f_e$ for 2 epochs with a learning rate of $2\times10^{-5}$, batch size 1, and 16-step gradient accumulation.
We use AdamW ($\epsilon=1\times10^{-8}$, weight decay $1\times10^{-2}$) with gradient norm clipping at 1.
All parameters are updated while regularizing the UltraEdit-modified subset $\theta_{edit}$ via $\mathcal{L}_{reg}$.

\begin{table*}[t]
\centering
\vspace{-0.1cm}
\caption{Security effectiveness (left) and functional correctness (right). We report Security Ratio (SR) with 95\% confidence intervals across CWE scenarios (in subscript) and Pass@1. Parentheses indicate relative change of \toolname{} compared to the respective baseline.}
\label{tab:imp_results}
\vspace{-0.3cm}
\tiny
\setlength{\tabcolsep}{2pt}
\renewcommand{\arraystretch}{1.2}
\begin{tabular}{llccc|ccc}
\toprule
\multirow{2}{*}{\textbf{Model}} & \multirow{2}{*}{\textbf{Method}} &
\multicolumn{3}{c|}{\textbf{Security Ratio (\%) [95\% CI]}} &
\multicolumn{3}{c}{\textbf{Pass@1 (\%)}} \\
\cmidrule(lr){3-5}\cmidrule(lr){6-8}
& & $T{=}0.1$ & $T{=}0.4$ & $T{=}0.8$ & $T{=}0.1$ & $T{=}0.4$ & $T{=}0.8$ \\
\midrule

\multirow{3}{*}{\shortstack[l]{\textbf{Llama3.1-8B}}}
& Vanilla     & 68.6{$_{[44.0,93.2]}$} ({\color{green!60!black}$\uparrow$2.92\%})  & 68.8{$_{[51.3,86.3]}$} ({\color{green!60!black}$\uparrow$1.31\%})  & 65.8{$_{[50.4,81.1]}$} ({\color{green!60!black}$\uparrow$5.32\%})  & 38.1 ({\color{green!60!black}$\uparrow$19.95\%}) & 36.7 ({\color{green!60!black}$\uparrow$24.52\%}) & 31.9 ({\color{green!60!black}$\uparrow$31.66\%}) \\
& CoSec       & 73.0{$_{[56.2,89.9]}$} ($\downarrow$3.29\%)  & 77.4{$_{[61.7,93.2]}$} ($\downarrow$9.95\%)  & 73.9{$_{[59.5,88.2]}$} ($\downarrow$6.22\%)  & 35.4 ({\color{green!60!black}$\uparrow$29.10\%}) & 36.1 ({\color{green!60!black}$\uparrow$26.59\%}) & 33.8 ({\color{green!60!black}$\uparrow$24.26\%}) \\
& \toolname{} & 70.6{$_{[45.8,91.7]}$} & 69.7{$_{[44.7,86.4]}$} & 69.3{$_{[51.2,85.4]}$} & 45.7 & 45.7 & 42.0 \\
\midrule

\multirow{3}{*}{\shortstack[l]{\textbf{Llama3.2-3B}}}
& Vanilla     & 61.3{$_{[38.6,83.9]}$} ({\color{green!60!black}$\uparrow$33.12\%}) & 62.2{$_{[44.3,80.1]}$} ({\color{green!60!black}$\uparrow$24.12\%}) & 64.8{$_{[51.6,78.0]}$} ({\color{green!60!black}$\uparrow$11.27\%}) & 26.7 ({\color{green!60!black}$\uparrow$10.86\%}) & 26.7 ({\color{green!60!black}$\uparrow$14.61\%}) & 24.0 ({\color{green!60!black}$\uparrow$17.50\%}) \\
& CoSec       & 65.7{$_{[49.2,82.1]}$} ({\color{green!60!black}$\uparrow$24.20\%}) & 67.4{$_{[52.8,81.9]}$} ({\color{green!60!black}$\uparrow$14.54\%}) & 63.6{$_{[48.4,78.7]}$} ({\color{green!60!black}$\uparrow$13.36\%}) & 24.1 ({\color{green!60!black}$\uparrow$22.82\%}) & 22.8 ({\color{green!60!black}$\uparrow$34.21\%}) & 19.8 ({\color{green!60!black}$\uparrow$42.42\%}) \\
& \toolname{} & 81.6{$_{[64.2,99.0]}$} & 77.2{$_{[62.6,91.8]}$} & 72.1{$_{[59.1,85.1]}$} & 29.6 & 30.6 & 28.2 \\
\midrule

\multirow{3}{*}{\shortstack[l]{\textbf{DeepSeek-6.7B}}}
& Vanilla     & 77.8{$_{[56.5,99.1]}$} ($\downarrow$7.20\%)  & 71.3{$_{[51.6,91.0]}$} ({\color{green!60!black}$\uparrow$2.66\%})  & 70.6{$_{[53.6,87.6]}$} ({\color{green!60!black}$\uparrow$4.53\%})  & 46.7 ({\color{green!60!black}$\uparrow$26.55\%}) & 43.6 ({\color{green!60!black}$\uparrow$31.88\%}) & 39.0 ({\color{green!60!black}$\uparrow$41.03\%}) \\
& CoSec       & 77.8{$_{[60.8,94.9]}$} ($\downarrow$7.20\%)  & 73.5{$_{[55.1,91.9]}$} ($\downarrow$0.41\%)  & 74.5{$_{[58.4,90.6]}$} ($\downarrow$0.94\%)  & 38.8 ({\color{green!60!black}$\uparrow$52.32\%}) & 39.2 ({\color{green!60!black}$\uparrow$46.68\%}) & 35.2 ({\color{green!60!black}$\uparrow$56.25\%}) \\
& \toolname{} & 72.2{$_{[49.3,95.1]}$} & 73.2{$_{[51.9,94.6]}$} & 73.8{$_{[55.5,92.2]}$} & 59.1 & 57.5 & 55.0 \\
\midrule

\multirow{3}{*}{\shortstack[l]{\textbf{Qwen2.5-C-7B}}}
& Vanilla     & 76.4{$_{[55.3,97.5]}$} ({\color{green!60!black}$\uparrow$11.65\%}) & 79.5{$_{[65.6,93.4]}$} ({\color{green!60!black}$\uparrow$7.92\%})  & 80.5{$_{[68.7,92.2]}$} ({\color{green!60!black}$\uparrow$2.11\%})  & 82.7 ($\downarrow$10.40\%) & 82.2 ($\downarrow$9.61\%)  & 79.3 ($\downarrow$9.21\%)  \\
& CoSec       & 75.7{$_{[62.9,88.6]}$} ({\color{green!60!black}$\uparrow$12.68\%}) & 76.6{$_{[64.8,88.3]}$} ({\color{green!60!black}$\uparrow$12.01\%}) & 76.7{$_{[64.7,88.7]}$} ({\color{green!60!black}$\uparrow$7.17\%})  & 75.3 ($\downarrow$1.59\%)  & 75.2 ($\downarrow$1.20\%)  & 75.1 ($\downarrow$4.13\%)  \\
& \toolname{} & 85.3{$_{[62.0,99.7]}$} & 85.8{$_{[70.3,99.6]}$} & 82.2{$_{[70.6,96.9]}$} & 74.1 & 74.3 & 72.0 \\
\midrule

\multirow{3}{*}{\shortstack[l]{\textbf{CodeGen-6B}}}
& Vanilla     & 67.0{$_{[45.3,88.8]}$} ({\color{green!60!black}$\uparrow$8.81\%})  & 65.4{$_{[47.4,83.4]}$} ({\color{green!60!black}$\uparrow$14.68\%}) & 68.0{$_{[54.1,81.9]}$} ({\color{green!60!black}$\uparrow$6.62\%})  & 18.6 ({\color{green!60!black}$\uparrow$35.48\%}) & 18.1 ({\color{green!60!black}$\uparrow$35.91\%}) & 14.0 ({\color{green!60!black}$\uparrow$50.00\%}) \\
& CoSec       & 71.3{$_{[55.9,86.7]}$} ({\color{green!60!black}$\uparrow$2.24\%})  & 69.8{$_{[55.0,84.7]}$} ({\color{green!60!black}$\uparrow$7.45\%})  & 71.6{$_{[56.7,86.5]}$} ({\color{green!60!black}$\uparrow$1.26\%})  & 12.4 ({\color{green!60!black}$\uparrow$103.23\%})& 12.3 ({\color{green!60!black}$\uparrow$100.00\%})& 10.6 ({\color{green!60!black}$\uparrow$98.11\%}) \\
& \toolname{} & 72.9{$_{[52.0,93.8]}$} & 75.0{$_{[57.2,92.8]}$} & 72.5{$_{[57.0,88.1]}$} & 25.2 & 24.6 & 21.0 \\
\midrule

\multirow{3}{*}{\shortstack[l]{\textbf{Qwen2.5-C-3B}}}
& Vanilla     & 80.3{$_{[61.6,99.0]}$} ($\downarrow$0.87\%)  & 75.0{$_{[59.1,91.0]}$} ({\color{green!60!black}$\uparrow$3.33\%})  & 67.1{$_{[53.8,80.3]}$} ({\color{green!60!black}$\uparrow$11.18\%}) & 59.6 ({\color{green!60!black}$\uparrow$19.46\%}) & 58.2 ({\color{green!60!black}$\uparrow$19.76\%}) & 53.3 ({\color{green!60!black}$\uparrow$24.58\%}) \\
& CoSec       & 64.9{$_{[53.1,76.6]}$} ({\color{green!60!black}$\uparrow$22.65\%}) & 62.0{$_{[50.0,74.1]}$} ({\color{green!60!black}$\uparrow$25.00\%}) & 65.8{$_{[55.3,76.3]}$} ({\color{green!60!black}$\uparrow$13.37\%}) & 43.1 ({\color{green!60!black}$\uparrow$65.20\%}) & 42.7 ({\color{green!60!black}$\uparrow$63.23\%}) & 42.5 ({\color{green!60!black}$\uparrow$56.24\%}) \\
& \toolname{} & 79.6{$_{[59.8,99.4]}$} & 77.5{$_{[58.8,96.3]}$} & 74.6{$_{[59.5,89.8]}$} & 71.2 & 69.7 & 66.4 \\
\midrule

\multirow{3}{*}{\shortstack[l]{\textbf{StarCoder-3B}}}
& Vanilla     & 63.9{$_{[40.1,87.7]}$} ({\color{green!60!black}$\uparrow$23.47\%}) & 66.1{$_{[47.1,85.1]}$} ({\color{green!60!black}$\uparrow$19.06\%}) & 65.8{$_{[51.2,80.4]}$} ({\color{green!60!black}$\uparrow$13.37\%}) & 16.2 ({\color{green!60!black}$\uparrow$81.48\%})  & 14.2 ({\color{green!60!black}$\uparrow$104.93\%}) & 11.2 ({\color{green!60!black}$\uparrow$139.29\%}) \\
& CoSec       & 62.6{$_{[52.2,73.1]}$} ({\color{green!60!black}$\uparrow$26.04\%}) & 63.3{$_{[52.9,73.7]}$} ({\color{green!60!black}$\uparrow$24.33\%}) & 61.9{$_{[51.2,72.6]}$} ({\color{green!60!black}$\uparrow$20.52\%}) & 13.5 ({\color{green!60!black}$\uparrow$117.78\%}) & 14.4 ({\color{green!60!black}$\uparrow$102.08\%}) & 12.7 ({\color{green!60!black}$\uparrow$111.02\%}) \\
& \toolname{} & 78.9{$_{[58.6,99.2]}$} & 78.7{$_{[61.7,95.7]}$} & 74.6{$_{[58.3,90.9]}$} & 29.4 & 29.1 & 26.8 \\
\midrule

\multirow{3}{*}{\shortstack[l]{\textbf{StarCoder-7B}}}
& Vanilla     & 72.5{$_{[52.4,92.6]}$} ({\color{green!60!black}$\uparrow$16.83\%}) & 69.5{$_{[51.9,87.0]}$} ({\color{green!60!black}$\uparrow$19.71\%}) & 70.0{$_{[56.6,83.3]}$} ({\color{green!60!black}$\uparrow$16.71\%}) & 22.2 ({\color{green!60!black}$\uparrow$65.77\%})  & 19.2 ({\color{green!60!black}$\uparrow$91.15\%})  & 16.3 ({\color{green!60!black}$\uparrow$109.82\%}) \\
& CoSec       & 71.2{$_{[55.0,87.4]}$} ({\color{green!60!black}$\uparrow$18.96\%}) & 67.9{$_{[51.9,84.0]}$} ({\color{green!60!black}$\uparrow$22.53\%}) & 73.1{$_{[61.0,85.2]}$} ({\color{green!60!black}$\uparrow$11.76\%}) & 17.3 ({\color{green!60!black}$\uparrow$112.72\%}) & 17.7 ({\color{green!60!black}$\uparrow$107.34\%}) & 17.2 ({\color{green!60!black}$\uparrow$98.84\%})  \\
& \toolname{} & 84.7{$_{[67.1,102.4]}$} & 83.2{$_{[66.1,100.4]}$} & 81.7{$_{[67.3,96.1]}$} & 36.8 & 36.7 & 34.2 \\
\bottomrule
\end{tabular}
\vspace{-0.1cm}
\end{table*}

\begin{wraptable}{r}{0.54\textwidth}
\centering
\caption{Statistical analysis of \toolname{} improvements over baselines across 8 LLMs, using Wilcoxon signed-rank test ($p$-value) and Cliff's $\delta$ effect size. Effect size: $|\delta|<0.147$ negligible, $[0.147,0.330)$ small, $[0.330,0.474)$ medium, $\geq0.474$ large.}
\label{tab:safeedit_stat}
\tiny
\setlength{\tabcolsep}{0.5pt}
\begin{tabular}{llllccc}
\toprule
\textbf{Metric} & \textbf{Comparison} & \textbf{Stat.} & $T{=}0.1$ & $T{=}0.4$ & $T{=}0.8$ \\
\midrule
\multirow{4}{*}{\textbf{SR}}
& \multirow{2}{*}{\toolname{} vs Vanilla}
  & $p$             & 0.027 & 0.004 & 0.004 \\
& & Cliff's $\delta$ & 0.594 (large) & 0.703 (large) & 0.750 (large) \\
\cmidrule(lr){2-6}
& \multirow{2}{*}{\toolname{} vs CoSec}
  & $p$             & 0.039 & 0.027 & 0.027 \\
& & Cliff's $\delta$ & 0.656 (large) & 0.688 (large) & 0.438 (medium) \\
\midrule
\multirow{4}{*}{\textbf{Pass@1}}
& \multirow{2}{*}{\toolname{} vs Vanilla}
  & $p$             & 0.027 & 0.020 & 0.020 \\
& & Cliff's $\delta$ & 0.250 (small) & 0.297 (small) & 0.344 (medium) \\
\cmidrule(lr){2-6}
& \multirow{2}{*}{\toolname{} vs CoSec}
  & $p$             & 0.008 & 0.008 & 0.008 \\
& & Cliff's $\delta$ & 0.406 (medium) & 0.406 (medium) & 0.375 (medium) \\
\bottomrule
\end{tabular}
\vspace{-0.2cm}
\end{wraptable}

\vspace{-0.2cm}
\subsubsection{Performance Comparison with Baselines}
We follow the empirical study setup and adopt the same six target LLMs under three decoding temperatures ($T \in \{0.1, 0.4, 0.8\}$), measuring both security effectiveness (Security Ratio via CodeQL) and functional correctness (Pass@1 on HumanEval).
To further assess the generalizability of \toolname{}, we additionally introduce two LLMs: StarCoder-3B and StarCoder-7B~\cite{Li2023}.
We compare \toolname{} against the vanilla model and CoSec to assess whether it improves the security--correctness trade-off over the unhardened baseline and the state-of-the-art inference-time hardening method.

Table~\ref{tab:imp_results} reports the security effectiveness and functional correctness of \toolname{} across eight target LLMs under three decoding temperatures.
Overall, \toolname{} consistently restores functional utility while largely preserving the security behavior injected by UltraEdit.
\textbf{(1) \textit{Compared to CoSec, \toolname{} improves both security and correctness}.}
Across all models, \toolname{} achieves higher security ratios than CoSec, with mean relative gains of +12.04\%, +11.94\%, and +7.54\% at $T=0.1$, $0.4$, and $0.8$, respectively.
For example, on StarCoder-3B, \toolname{} increases the security ratio by 26.04\% at $T=0.1$.
More importantly, \toolname{} obtains substantially stronger functional correctness, improving average Pass@1 over CoSec by +62.70\%, +59.87\%, and +60.38\% under the three temperatures.
\textbf{(2) \textit{Compared to UltraEdit, \toolname{} mitigates functional regressions with minimal security loss}.}
Relative to UltraEdit, \toolname{} largely preserves security effectiveness, with average absolute SR changes of +7.44, -0.28, and -3.48 percentage points at $T=0.1$, $0.4$, and $0.8$, respectively.
Meanwhile, it substantially restores functional correctness, improving Pass@1 by +11.73, +13.70, and +15.50 percentage points under the three temperatures.
This shows that edit-aware regularization effectively prevents functional recovery from overwriting the injected security behavior, while substantially reducing the correctness degradation caused by direct security editing.
Table~\ref{tab:safeedit_stat} further reports Wilcoxon signed-rank tests~\cite{Wilcoxon1992} and Cliff's $\delta$~\cite{Cliff1993} over the eight LLMs at each temperature.
\toolname{} significantly improves SR over CoSec at all temperatures ($p \le 0.039$, $\delta=0.438$--$0.688$) and Pass@1 over CoSec ($p=0.008$, $\delta=0.375$--$0.406$).
It also significantly improves over Vanilla in both SR and Pass@1.
These results indicate that \toolname{} provides statistically reliable improvements in both security and functional correctness.

\begin{table*}[b]
\centering
\caption{Performance on CodeGuard+~\cite{Fu2024}. P@1: Pass@1 (functional correctness); S-Rate: Security ratio; S@1P: Secure@1-Pass (secure among correct completions); SP@1: Secure-Pass@1 (jointly secure and correct). Parentheses show relative change of \toolname{} over each baseline. Best per model in \textbf{bold}. Model abbreviations: L3.1=LLaMA3.1-8B, L3.2=LLaMA3.2-3B, DS=DeepSeek-6.7B, Q2.5-7B=Qwen2.5-Coder-7B, Q2.5-3B=Qwen2.5-Coder-3B, CG=CodeGen-6B, SC=StarCoder.}
\label{tab:codeguardplus}
\vspace{-0.3cm}
\tiny
\setlength{\tabcolsep}{0pt}
\begin{tabular}{llcccc||llcccc}
\toprule
\textbf{Model} & \textbf{Met.} & \textbf{P@1} & \textbf{S-Rate} & \textbf{S@1P} & \textbf{SP@1} &
\textbf{Model} & \textbf{Met.} & \textbf{P@1} & \textbf{S-Rate} & \textbf{S@1P} & \textbf{SP@1} \\
\midrule

\multirow{3}{*}{\textbf{L3.1-8B}}
& Van. & 79.03 ({\color{green!60!black}$\uparrow$7.49\%})  & 60.04 ({\color{green!60!black}$\uparrow$2.47\%})  & 54.16 ({\color{green!60!black}$\uparrow$15.01\%}) & 45.92 ({\color{green!60!black}$\uparrow$15.22\%}) &
\multirow{3}{*}{\textbf{CG-6B}}
& Van. & 59.61 ({\color{green!60!black}$\uparrow$8.96\%})  & 57.63 ({\color{green!60!black}$\uparrow$7.46\%})  & 49.11 ({\color{green!60!black}$\uparrow$0.31\%})  & 33.50 ({\color{green!60!black}$\uparrow$13.61\%}) \\
& CoS. & 46.80 ({\color{green!60!black}$\uparrow$81.52\%}) & 59.29 ({\color{green!60!black}$\uparrow$3.76\%})  & 50.32 ({\color{green!60!black}$\uparrow$23.79\%}) & 27.28 ({\color{green!60!black}$\uparrow$93.95\%}) &
& CoS. & 45.05 ({\color{green!60!black}$\uparrow$44.17\%}) & 59.79 ({\color{green!60!black}$\uparrow$3.58\%})  & 45.86 ({\color{green!60!black}$\uparrow$7.41\%})  & 24.76 ({\color{green!60!black}$\uparrow$53.72\%}) \\
& \toolname{} & \textbf{84.95} & \textbf{61.52} & \textbf{62.29} & \textbf{52.91} &
& \toolname{} & \textbf{64.95} & \textbf{61.93} & \textbf{49.26} & \textbf{38.06} \\
\midrule

\multirow{3}{*}{\textbf{L3.2-3B}}
& Van. & \textbf{75.63} ($\downarrow$11.94\%) & 55.78 ({\color{green!60!black}$\uparrow$19.34\%}) & 53.94 ({\color{green!60!black}$\uparrow$5.38\%})  & 40.29 ({\color{green!60!black}$\uparrow$7.00\%})  &
\multirow{3}{*}{\textbf{DS-6.7B}}
& Van. & 82.62 ({\color{green!60!black}$\uparrow$7.06\%})  & 57.86 ({\color{green!60!black}$\uparrow$9.25\%})  & 56.92 ({\color{green!60!black}$\uparrow$3.20\%})  & 47.67 ({\color{green!60!black}$\uparrow$19.34\%}) \\
& CoS. & 40.68 ({\color{green!60!black}$\uparrow$63.72\%}) & 59.92 ({\color{green!60!black}$\uparrow$11.10\%}) & 42.44 ({\color{green!60!black}$\uparrow$33.93\%}) & 21.26 ({\color{green!60!black}$\uparrow$102.78\%}) &
& CoS. & 82.04 ({\color{green!60!black}$\uparrow$7.81\%})  & 58.62 ({\color{green!60!black}$\uparrow$7.83\%})  & 57.53 ({\color{green!60!black}$\uparrow$2.10\%})  & 48.64 ({\color{green!60!black}$\uparrow$16.96\%}) \\
& \toolname{} & 66.60 & \textbf{66.57} & \textbf{56.84} & \textbf{43.11} &
& \toolname{} & \textbf{88.45} & \textbf{63.21} & \textbf{58.74} & \textbf{56.89} \\
\midrule

\multirow{3}{*}{\textbf{Q2.5-3B}}
& Van. & 80.00 ({\color{green!60!black}$\uparrow$5.10\%})  & 60.54 ({\color{green!60!black}$\uparrow$5.68\%})  & 56.70 ({\color{green!60!black}$\uparrow$6.10\%})  & 45.24 ({\color{green!60!black}$\uparrow$18.26\%}) &
\multirow{3}{*}{\textbf{Q2.5-7B}}
& Van. & 83.59 ({\color{green!60!black}$\uparrow$2.44\%})  & 65.99 ({\color{green!60!black}$\uparrow$2.24\%})  & 65.25 ({\color{green!60!black}$\uparrow$1.27\%})  & 55.63 ({\color{green!60!black}$\uparrow$5.41\%})  \\
& CoS. & 75.53 ({\color{green!60!black}$\uparrow$11.32\%}) & 59.59 ({\color{green!60!black}$\uparrow$7.37\%})  & 57.60 ({\color{green!60!black}$\uparrow$4.44\%})  & 43.50 ({\color{green!60!black}$\uparrow$22.99\%}) &
& CoS. & 83.88 ({\color{green!60!black}$\uparrow$2.09\%})  & 62.49 ({\color{green!60!black}$\uparrow$7.97\%})  & 60.94 ({\color{green!60!black}$\uparrow$8.43\%})  & 52.23 ({\color{green!60!black}$\uparrow$12.27\%}) \\
& \toolname{} & \textbf{84.08} & \textbf{63.98} & \textbf{60.16} & \textbf{53.50} &
& \toolname{} & \textbf{85.63} & \textbf{67.47} & \textbf{66.08} & \textbf{58.64} \\
\midrule

\multirow{3}{*}{\textbf{SC-3B}}
& Van. & 69.13 ({\color{green!60!black}$\uparrow$9.40\%})  & 59.49 ({\color{green!60!black}$\uparrow$3.09\%})  & 54.31 ($\downarrow$1.91\%)           & 40.49 ({\color{green!60!black}$\uparrow$11.02\%}) &
\multirow{3}{*}{\textbf{SC-7B}}
& Van. & 78.64 ({\color{green!60!black}$\uparrow$7.53\%})  & 57.21 ({\color{green!60!black}$\uparrow$13.86\%}) & 55.35 ({\color{green!60!black}$\uparrow$11.02\%}) & 45.44 ({\color{green!60!black}$\uparrow$20.29\%}) \\
& CoS. & 60.39 ({\color{green!60!black}$\uparrow$25.24\%}) & 60.90 ({\color{green!60!black}$\uparrow$0.71\%})  & 53.42 ($\downarrow$0.28\%)           & 35.44 ({\color{green!60!black}$\uparrow$26.83\%}) &
& CoS. & 67.48 ({\color{green!60!black}$\uparrow$25.31\%}) & 58.61 ({\color{green!60!black}$\uparrow$11.14\%}) & 54.81 ({\color{green!60!black}$\uparrow$12.11\%}) & 39.42 ({\color{green!60!black}$\uparrow$38.66\%}) \\
& \toolname{} & \textbf{75.63} & \textbf{61.33} & 53.27         & \textbf{44.95} &
& \toolname{} & \textbf{84.56} & \textbf{65.14} & \textbf{61.45} & \textbf{54.66} \\
\bottomrule
\end{tabular}
\vspace{-0.2cm}
\end{table*}


\vspace{-0.2cm}
\subsubsection{Evaluation on CodeGuard+}
\label{sec:codeguardplus}
To further strengthen our evaluation, we conduct an additional experiment on CodeGuard+~\cite{Fu2024}, a recent benchmark that jointly evaluates the security and functional correctness of LLM-generated code.
CodeGuard+ contains 91 prompts covering 34 CWE categories in Python and C/C++, provides unit tests, and uses a CodeQL--SonarQube ensemble for vulnerability detection.
Following its protocol, a generated program is considered vulnerable if either analyzer reports a vulnerability, and secure only if both regard it as secure.
Thus, CodeGuard+ complements our original evaluation by broadening CWE coverage, adding unit-test-based correctness checks, and reducing reliance on CodeQL alone.
We evaluate the same models and hardening methods under the default decoding configuration.
We report four metrics: security ratio, pass@1, secure@1-Pass, and secure-pass@1.
The first two measure security and functional correctness separately, secure@1-Pass measures security among functionally correct generations, and secure-pass@1 measures whether a generation is both secure and correct.
We use secure-pass@1 as the primary evaluation metric because it best reflects the practical goal of secure code generation.

From Table~\ref{tab:codeguardplus}, in general \toolname{} achieves the best joint security--correctness performance across all evaluated models.
Compared with vanilla models, \toolname{} improves the average P@1 from 76.03 to 79.36, S-Rate from 59.32 to 63.89.
The gain is most consistent on SP@1, the strictest metric requiring generated code to be both secure and functionally correct: \toolname{} improves over vanilla on all eight models, with relative gains from 5.41\% to 20.29\%.
\toolname{} also consistently outperforms CoSec.
On average, it improves SP@1 from 36.57 to 50.34, and achieves higher P@1, S-Rate, and S@1P as well.
These results provide additional evidence that under CodeGuard+'s stricter CodeQL--SonarQube ensemble and unit-test-based functional validation, \toolname{} still improves the probability of generating code that is both secure and correct.

\myfinding{4}{
\toolname{} restores functional correctness while largely preserving UltraEdit's injected security behavior.
Compared to CoSec, it improves SR by +12.04\%/+11.94\%/+7.54\% and Pass@1 by +62.70\%/+59.87\%/+60.38\% at $T=0.1/0.4/0.8$.
CodeGuard+ further confirms these gains under joint security--correctness validation.
}

\color{black}

\color{black}

\vspace{-0.3cm}
\section{Discussion}

\subsection{Complementary Advantages}

\begin{table*}[h]
\centering
\caption{Complementary advantages of \toolname{} and CoSec. We report SR (left) and Pass@1 (right).}
\label{tab:hybrid_results}
\vspace{-0.3cm}
\tiny
\setlength{\tabcolsep}{3.5pt}
\begin{tabular}{l l c c c @{\hspace{8pt}}|@{\hspace{8pt}} c c c}
\toprule
\multirow{2}{*}{\textbf{Model}} & \multirow{2}{*}{\textbf{Method}} &
\multicolumn{3}{c}{\textbf{Security Ratio (\%)}} &
\multicolumn{3}{c}{\textbf{Pass@1 (\%)}} \\
\cmidrule(lr){3-8}
& & \textbf{$T=0.1$} & \textbf{$T=0.4$} & \textbf{$T=0.8$}
& \textbf{$T=0.1$} & \textbf{$T=0.4$} & \textbf{$T=0.8$} \\
\midrule

\multirow{3}{*}{\makecell[l]{LLaMA\\3.1-8B}}
& CoSec
& 73.0 ($\downarrow$ 3.01\%)  & 77.4 ($\downarrow$ 7.11\%)  & 73.9 ({\color{green!60!black}$\uparrow$ 0.41\%}) 
& 35.4 ({\color{green!60!black}$\uparrow$ 31.64\%})  & 36.1 ({\color{green!60!black}$\uparrow$ 24.93\%})  & 33.8 ({\color{green!60!black}$\uparrow$ 28.40\%})  \\
& \toolname{}
& 70.6 ({\color{green!60!black}$\uparrow$ 0.28\%}) & 69.7 ({\color{green!60!black}$\uparrow$ 3.16\%}) & 69.3 ({\color{green!60!black}$\uparrow$ 7.07\%})
& 45.7 ({\color{green!60!black}$\uparrow$ 1.97\%}) & 45.7 ($\downarrow$ 1.31\%)  & 42.0 ({\color{green!60!black}$\uparrow$ 3.33\%}) \\
& \toolname{}+CoSec
& 70.8  & 71.9   & 74.2  
& 46.6   & 45.1  & 43.4  \\
\midrule

\multirow{3}{*}{\makecell[l]{LLaMA\\3.2-3B}}
& CoSec
& 65.7 ({\color{green!60!black}$\uparrow$ 21.77\%})  & 67.4 ({\color{green!60!black}$\uparrow$ 18.69\%}) & 63.6 ({\color{green!60!black}$\uparrow$ 24.84\%}) 
& 24.1 ({\color{green!60!black}$\uparrow$ 47.30\%})  & 22.8 ({\color{green!60!black}$\uparrow$ 44.30\%}) & 19.8 ({\color{green!60!black}$\uparrow$ 46.46\%})  \\
& \toolname{}
& 81.6 ($\downarrow$ 1.96\%)  & 77.2 ({\color{green!60!black}$\uparrow$ 3.63\%}) & 72.1 ({\color{green!60!black}$\uparrow$ 10.12\%})
& 29.6 ({\color{green!60!black}$\uparrow$ 19.93\%}) & 30.6 ({\color{green!60!black}$\uparrow$ 7.52\%})  & 28.2 ({\color{green!60!black}$\uparrow$ 2.94\%}) \\
& \toolname{}+CoSec
& 80.0   & 80.0    & 79.4
& 35.5  & 32.9   & 29.0  \\

\bottomrule
\end{tabular}
\vspace{-0.05cm}
\end{table*}

Although \toolname{} and CoSec adopt fundamentally different hardening mechanisms, they provide complementary strengths: \toolname{} improves model-level security and functional correctness via edit-aware refinement, while CoSec provides runtime adaptability through inference-time token steering.
To validate this complementarity, we apply CoSec on top of \toolname{}-refined models and compare three settings on LLaMA3.1-8B and LLaMA3.2-3B: (i) CoSec alone, (ii) \toolname{} alone, and (iii) \toolname{}+CoSec.
Table~\ref{tab:hybrid_results} reports the SR and Pass@1 when combining \toolname{} with CoSec on two representative backbones. 
Overall, the results confirm that the combined setting can yield strictly stronger security than either method alone under stochastic decoding.
On LLaMA3.2-3B at $T=0.8$, \toolname{}+CoSec reaches an SR of 79.4\%, substantially higher than CoSec (63.6\%) and also higher than \toolname{} alone (72.1\%), showing that inference-time steering and model-level refinement reinforce each other rather than overlapping.
Importantly, this security boost does not sacrifice utility: across both backbones, \toolname{}+CoSec consistently achieves higher Pass@1 than either \toolname{} or CoSec alone across almost all decoding temperatures.


\vspace{-0.3cm}
\subsection{Generalization of \toolname{} to Unseen CWEs}
\label{sec:safeedit-unseen}

\begin{wraptable}{r}{0.45\textwidth}
\centering
\tiny
\vspace{-0.2cm}
\caption{Security Ratio (\%) on unseen CWEs for \toolname{}, Vanilla, and UltraEdit.}
\label{tab:safeedit_unseen}
\vspace{-0.2cm}
\setlength{\tabcolsep}{4pt}
\renewcommand{\arraystretch}{1.05}
\begin{tabular}{llcccc}
\toprule
\textbf{Model} & \textbf{Method} & $T{=}0.1$ & $T{=}0.4$ & $T{=}0.8$ & \textbf{Avg.} \\
\midrule
\multirow{3}{*}{LLaMA3.2-3B}
& Vanilla   & 45.5 & 49.4 & 53.1 & 49.3 \\
& UltraEdit & 44.9 & 49.9 & 51.7 & 48.8 \\
& \toolname{}  & \textbf{51.2} & \textbf{54.6} & \textbf{58.5} & \textbf{54.8} \\
\midrule
\multirow{3}{*}{LLaMA3.1-8B}
& Vanilla   & 51.0 & \textbf{54.9} & 54.9 & 53.6 \\
& UltraEdit & 51.1 & \textbf{54.9} & 55.2 & 53.7 \\
& \toolname{}  & \textbf{52.7} & 53.9 & \textbf{56.3} & \textbf{54.3} \\
\bottomrule
\end{tabular}
\vspace{-0.2cm}
\end{wraptable}

Since \toolname{} refines UltraEdit to recover functional correctness, we further examine whether it weakens security generalization on unseen CWEs.
Table~\ref{tab:safeedit_unseen} reports the results on LLaMA3.1-8B and LLaMA3.2-3B.
\toolname{} preserves and often improves UltraEdit's unseen-CWE security: on LLaMA3.2-3B, it raises average SR from 49.3\%/48.8\% for Vanilla/UltraEdit to 54.8\%, with consistent gains across all temperatures; on LLaMA3.1-8B, it also achieves the best average SR, improving from 53.6\%/53.7\% to 54.3\%.
These results indicate that \toolname{}'s functional recovery does not erase the transferable security behavior introduced by editing, and can even improve unseen-CWE generalization in some settings.

\vspace{-0.35cm}
\subsection{Cross-language Functional Correctness.}

\begin{wraptable}{r}{0.45\textwidth}
\centering
\vspace{-0.45cm}
\caption{MultiPL-E Pass@1 (\%) at $T{=}0.4$, averaged over C++, JavaScript, and TypeScript. Best per model in \textbf{bold}.}
\label{tab:multiple_e}
\vspace{-0.3cm}
\tiny
\setlength{\tabcolsep}{1pt}
\begin{tabular}{llc||llc}
\toprule
\textbf{Model} & \textbf{Met.} & $T{=}0.4$ &
\textbf{Model} & \textbf{Met.} & $T{=}0.4$ \\
\midrule
\multirow{3}{*}{\textbf{L3.1-8B}}
& Van.        & 33.85 ({\color{green!60!black}$\uparrow$18.32\%}) &
\multirow{3}{*}{\textbf{CG-6B}}
& Van.        & 12.98 ({\color{green!60!black}$\uparrow$46.07\%}) \\
& CoSec       & 12.93 ({\color{green!60!black}$\uparrow$209.74\%}) &
& CoSec       & 6.49 ({\color{green!60!black}$\uparrow$192.14\%}) \\
& \toolname{} & \textbf{40.05} &
& \toolname{} & \textbf{18.96} \\
\midrule
\multirow{3}{*}{\textbf{L3.2-3B}}
& Van.        & 23.83 ({\color{green!60!black}$\uparrow$10.49\%}) &
\multirow{3}{*}{\textbf{DS-6.7B}}
& Van.        & 45.86 ({\color{green!60!black}$\uparrow$25.95\%}) \\
& CoSec       & 8.01 ({\color{green!60!black}$\uparrow$228.71\%}) &
& CoSec       & 36.95 ({\color{green!60!black}$\uparrow$56.32\%}) \\
& \toolname{} & \textbf{26.33} &
& \toolname{} & \textbf{57.76} \\
\midrule
\multirow{3}{*}{\textbf{Q2.5-3B}}
& Van.        & 48.87 ({\color{green!60!black}$\uparrow$26.01\%}) &
\multirow{3}{*}{\textbf{Q2.5-7B}}
& Van.        & 64.95 ({\color{green!60!black}$\uparrow$3.46\%}) \\
& CoSec       & 39.84 ({\color{green!60!black}$\uparrow$54.57\%}) &
& CoSec       & 56.22 ({\color{green!60!black}$\uparrow$19.53\%}) \\
& \toolname{} & \textbf{61.58} &
& \toolname{} & \textbf{67.20} \\
\midrule
\multirow{3}{*}{\textbf{SC-3B}}
& Van.        & 19.87 ({\color{green!60!black}$\uparrow$31.61\%}) &
\multirow{3}{*}{\textbf{SC-7B}}
& Van.        & 25.37 ({\color{green!60!black}$\uparrow$35.75\%}) \\
& CoSec       & 16.13 ({\color{green!60!black}$\uparrow$62.12\%}) &
& CoSec       & 19.75 ({\color{green!60!black}$\uparrow$74.38\%}) \\
& \toolname{} & \textbf{26.15} &
& \toolname{} & \textbf{34.44} \\
\bottomrule
\end{tabular}
\vspace{-0.2cm}
\end{wraptable}

To address the concern that HumanEval only involves Python, we further evaluate functional correctness on MultiPL-E~\cite{Cassano2023}.
We select C++, JavaScript, and TypeScript, and run the same models and hardening methods under the three decoding setups with $T \in \{0.1,0.4,0.8\}$.
Due to space limits, Table~\ref{tab:multiple_e} reports the average Pass@1 over the three languages at $T{=}0.4$, while full results are included in our replication package.
\toolname{} achieves the best Pass@1 on all eight models, improving over vanilla by 3.46\%--46.07\% and over CoSec by 19.53\%--228.71\%.
Overall, the MultiPL-E results further support the generality of \toolname{}'s functional preservation.

\vspace{-0.3cm}
\subsection{\toolname{} vs.\ Fine-tuning}

\begin{wraptable}{r}{0.52\textwidth}
\vspace{-0.1cm}
\centering
\caption{TradeOff score and MultiPL-E Pass@1 (avg.\ over JS/TS/C++) for \toolname{} vs.\ LoRA.}
\vspace{-0.2cm}
\label{tab:lora_comparison}
\tiny
\setlength{\tabcolsep}{3pt}
\renewcommand{\arraystretch}{1.2}
\begin{tabular}{llccc|ccc}
\toprule
\multirow{2}{*}{\textbf{Model}} & \multirow{2}{*}{\textbf{Method}} &
\multicolumn{3}{c|}{\textbf{TradeOff}} &
\multicolumn{3}{c}{\textbf{MultiPL-E P@1}} \\
\cmidrule(lr){3-5}\cmidrule(lr){6-8}
& & $T{=}0.1$ & $T{=}0.4$ & $T{=}0.8$ & $T{=}0.1$ & $T{=}0.4$ & $T{=}0.8$ \\
\midrule
\multirow{2}{*}{LLaMA3.1-8B}
& LoRA      & 52.9 & 52.7 & 48.8 & 34.92 & 33.71 & 27.76 \\
& \toolname{} & \textbf{55.4} & \textbf{55.2} & \textbf{52.3} & \textbf{41.48} & \textbf{40.05} & \textbf{37.29} \\
\midrule
\multirow{2}{*}{LLaMA3.2-3B}
& LoRA      & \textbf{45.4} & 42.8 & 37.1 & 25.62 & 23.27 & 17.91 \\
& \toolname{} & 43.4 & \textbf{43.8} & \textbf{40.6} & \textbf{26.58} & \textbf{26.33} & \textbf{23.75} \\
\bottomrule
\end{tabular}
\vspace{-0.2cm}
\end{wraptable}
\toolname{} addresses a problem orthogonal to standard fine-tuning: preserving \emph{security-injected knowledge} during functional recovery.
Its edit-aware regularization anchors $\theta_{edit}$, the exact parameter subset modified by UltraEdit, preventing functional tuning from overwriting injected security behavior, an objective absent from both the model editing and secure code generation literature.
Moreover, model editing operates under few-shot, low-cost constraints (50--100 seconds), whereas fine-tuning assumes large-scale security-annotated corpora and full retraining infrastructure.
Empirically, Table~\ref{tab:lora_comparison} reports both the TradeOff score and MultiPL-E Pass@1 averaged over JavaScript, TypeScript, and C++.
On LLaMA3.1-8B, \toolname{} consistently outperforms LoRA across all temperatures on both metrics (\eg TradeOff: 55.4 vs.\ 52.9 at $T = 0.1$; MultiPL-E Pass@1: 41.48 vs.\ 34.92).
On LLaMA3.2-3B, \toolname{} leads at higher temperatures (40.6 vs.\ 37.1 at $T=0.8$) and consistently achieves better cross-lingual generalization on MultiPL-E across all temperatures.

\vspace{-0.3cm}
\subsection{Sensitivity Analysis of Regularization Strength $\mu$}
\label{sec:sensitivity}

\begin{wraptable}{r}{0.45\textwidth}
\vspace{-0.2cm}
\centering
\caption{Sensitivity analysis of $\mu$ in \toolname{}.}
\vspace{-0.3cm}
\label{tab:mu_sensitivity}
\tiny
\setlength{\tabcolsep}{3pt}
\renewcommand{\arraystretch}{1.2}
\begin{tabular}{llcccc}
\toprule
\multirow{2}{*}{\textbf{Model}} &
\multirow{2}{*}{$\mu$} &
\multicolumn{2}{c}{\textbf{Security Ratio (\%)}} &
\multicolumn{2}{c}{\textbf{Pass@1 (\%)}} \\
\cmidrule(lr){3-4} \cmidrule(lr){5-6}
& & $T=0.1$ & $T=0.4$ & $T=0.1$ & $T=0.4$ \\
\midrule
\multirow{2}{*}{LLaMA3.1-8B}
& $10^{-3}$ & \textbf{70.6} & \textbf{69.7} & \textbf{45.7} & \textbf{45.7} \\
& $10^{-4}$ & 67.5 & 66.2 & 43.5 & 43.1 \\
\midrule
\multirow{2}{*}{Qwen2.5-Coder-7B}
& $10^{-3}$ & \textbf{85.3} & \textbf{85.8} & \textbf{74.1} & \textbf{74.3} \\
& $10^{-4}$ & 81.2 & 80.5 & 74.0 & 74.0 \\
\bottomrule
\end{tabular}
\end{wraptable}
To examine the sensitivity of \toolname{} to the regularization strength $\mu$, we compare $\mu \in \{10^{-3}, 10^{-4}\}$ on LLaMA3.1-8B and Qwen2.5-Coder-7B, reporting SR and Pass@1 at $T \in \{0.1, 0.4\}$.
Table~\ref{tab:mu_sensitivity} shows that $\mu=10^{-3}$ consistently outperforms $\mu=10^{-4}$ on both models.
On LLaMA3.1-8B, the stronger regularization improves SR by up to +3.5 points and Pass@1 by up to +2.6 points at $T=0.4$.
On Qwen2.5-Coder-7B, SR gains reach +5.3 points at $T=0.4$ while Pass@1 remains largely unaffected.
A smaller $\mu$ weakens the edit-aware regularization, allowing functional tuning to drift the edited parameters away from their post-edit values, which in turn degrades both security retention and functional correctness.

\color{black}

\subsection{Efficiency}
\label{sec:efficiency}

\begin{table*}[h]
\centering
\vspace{-0.1cm}
\caption{Efficiency comparison of various methods.
We report hardening (training) time and per-query inference latency (in seconds).
Lower is better for both metrics.}
\vspace{-0.3cm}
\label{tab:rq4_efficiency}
\tiny
\renewcommand{\arraystretch}{1.05}

\begin{tabular}{l rr rr rr rr rr rr rr rr rr rr}
\toprule
& \multicolumn{2}{c}{LLaMA3.1-8B}
& \multicolumn{2}{c}{LLaMA3.2-3B}
& \multicolumn{2}{c}{DeepSeek-6.7B}
& \multicolumn{2}{c}{Qwen2.5-Coder-7B}
& \multicolumn{2}{c}{CodeGen-6B}
& \multicolumn{2}{c}{Qwen2.5-Coder-3B} \\
\cmidrule(lr){2-3}\cmidrule(lr){4-5}\cmidrule(lr){6-7}\cmidrule(lr){8-9}\cmidrule(lr){10-11}\cmidrule(lr){12-13}
Method & Train & Infer & Train & Infer & Train & Infer & Train & Infer & Train & Infer & Train & Infer \\
\midrule
Vanilla   & ---     & 0.52 & ---     & 0.31 & ---     & 0.97 & ---     & 1.01 & ---     & 1.20 & ---     & 0.15 \\
CoSec     & 6,087.00 & 1.68 & 1,391.00 & 0.84 & 986.00  & 1.87 & 582.00  & 1.61 & 831.00  & 1.72 & 582.00  & 0.27 \\
UltraEdit & 94.52   & 0.44 & 56.88   & 0.29 & 83.96   & 1.19 & 86.45   & 0.98 & 102.80  & 1.29 & 56.51   & 0.15 \\
DINM      & 178.57  & 1.28 & 85.88   & 0.79 & 228.88  & 1.06 & 150.04  & 1.14 & 168.19  & 1.16 & 84.79   & 0.16 \\
DEFER*    & 1,735.49 & 0.81 & 833.00  & 0.36 & 1,531.47 & 0.99 & 1,608.50 & 0.98 & 1,804.28 & 1.32 & 861.69  & 0.17 \\
\toolname{}   &{25,719.00}  & {0.45}   & {10,210.00}  & {0.30}  & {176,055.00}  & {0.95}  & {96,387.00} & {0.94}  & {57,014.00} & {1.25}  & {30,785.00}  &  {0.15} \\

\bottomrule
\end{tabular}

\vspace{-0.05cm}
\end{table*}

We investigate the efficiency of different security hardening strategies, focusing on both offline hardening cost and online inference overhead.
Specifically, we report (i) hardening time (training or editing time required) and (ii) per-query inference latency for HumanEval, measured in seconds.
All methods are evaluated under the same hardware and decoding configuration as previous RQs.

Table~\ref{tab:rq4_efficiency} shows a clear separation between model editing methods and CoSec.
Model editing methods incur a one-time hardening cost but introduce little or no inference overhead, whereas CoSec shifts most of the cost to inference time.
UltraEdit achieves the best efficiency–effectiveness balance.
Among editing methods, UltraEdit achieves the best efficiency: hardening completes within 50--100 seconds even for 8B-scale models, and inference overhead is negligible (\eg LLaMA3.1-8B: 0.44s vs.\ 0.52s for vanilla), owing to its localized, lightweight updates on a small subset of projection matrices.
\toolname{} follows a different efficiency profile.
Because it performs an additional post-edit functional refinement stage after UltraEdit, its offline hardening time is substantially larger than UltraEdit alone (\eg 25,719s vs.\ 94.52s on LLaMA3.1-8B).
However, this cost is paid only once before deployment.
At inference time, \toolname{} behaves like a standard refined model and does not require an auxiliary model, routing module, or co-decoding.
As a result, its per-query latency remains comparable to vanilla and UltraEdit across all models (\eg 0.45s vs.\ 0.52s/0.44s on LLaMA3.1-8B).
CoSec avoids offline training on the target model, but consistently increases inference latency by $1.4$--$3.2{\times}$ across all evaluated models (\eg LLaMA3.1-8B: 1.68s vs.\ 0.52s for vanilla).
This persistent overhead stems from co-decoding with an auxiliary model at every inference step, which remains unavoidable regardless of auxiliary model size.
Therefore, \toolname{} is more suitable when an offline refinement budget is acceptable and low-latency deployment is important, whereas CoSec is preferable when parameter access or offline model updating is unavailable.


\vspace{-0.3cm}
\subsection{Editing Depth}


We conduct a layer-wise ablation by injecting security edits at three depths: close to the input (layer-10), moderately deep (layer-20), and at the default setting used in prior experiments.
Due to space constraints, we focus on LLaMA3.1-8B and CodeGen-6B, which have comparable model depths.
Results are reported under three decoding temperatures.


\begin{figure}[h]
    \centering
    \includegraphics[width=0.95\linewidth]{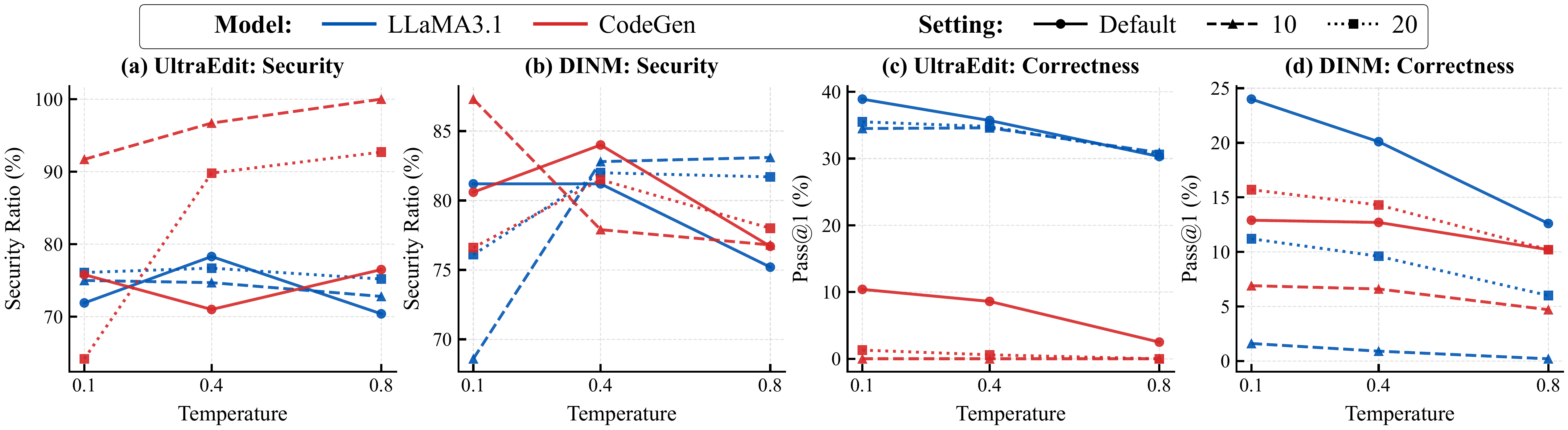}
    \vspace{-0.3cm}
    \caption{Impact of editing depth on security effectiveness and functional correctness for UltraEdit and DINM.
  }
  \vspace{-0.15cm}
    \label{fig:rq5_depth}
\end{figure}

Figure~\ref{fig:rq5_depth} reveals two consistent patterns.
Security effectiveness is sensitive to injection depth, and the optimal depth varies across methods.
UltraEdit benefits from earlier injection, with layer-10 achieving the highest SR on CodeGen-6B across temperatures.
DINM, however, does not consistently benefit from early-layer edits; on LLaMA3.1-8B, moving edits toward layer-10 even degrades security at low temperatures, suggesting that neuron-level updates require deeper contextual representations to be effective.
Functional correctness degrades markedly with earlier-layer edits for both methods.
DINM nearly collapses Pass@1 at layer-10 across both models, indicating severe disruption to foundational synthesis representations.
UltraEdit remains relatively stable but still incurs a noticeable correctness drop at early layers.
Overall, intermediate or later layers strike a better balance, supporting our design choice of injecting security knowledge near the output layers.

\subsection{Editing Suffix Length}

To investigate how the length of the target suffix $L$ affects the effectiveness, we ablate the target suffix length $L \in \{10, 50, 100\}$ on UltraEdit and DINM across LLaMA3.1-8B and Qwen2.5-Coder-7B under three decoding temperatures, where $L=50$ is the default.


\begin{figure}[h]
    \centering
    \includegraphics[width=0.95\linewidth]{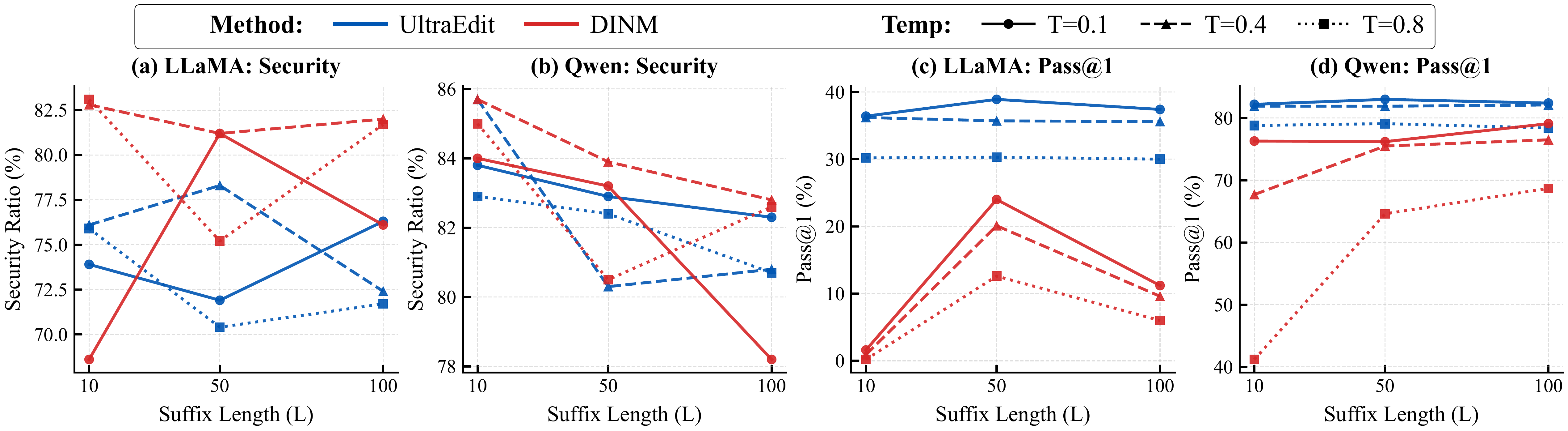}
    \vspace{-0.3cm}
    \caption{Effect of target suffix length $L$ on security effectiveness and functional correctness.}
        \vspace{-0.1cm}
    \label{fig:rq6}
\end{figure}

Figure~\ref{fig:rq6} shows that security effectiveness is sensitive to the target suffix length $L$, and the optimal $L$ depends on the editing method and model.
Shorter suffixes (\eg $L=10$) can sharpen security-relevant updates, particularly for DINM on Qwen2.5-Coder-7B, yet no single $L$ dominates consistently on LLaMA3.1-8B.
Functional correctness exhibits a clearer trend: DINM degrades substantially under short suffixes, as partial supervision weakens parameter update grounding and overemphasizes superficial security cues; Pass@1 nearly collapses at $L=10$ on LLaMA3.1-8B under low-temperature decoding.
UltraEdit, by contrast, remains robust across all $L$ values owing to its localized, lightweight update mechanism.


\subsection{Knowledge Injection Location}
To study how the location of parameter modification affects security editing, we conduct a targeted ablation on where security knowledge is injected within the transformer architecture.
Modern decoder-only LLMs (\eg LLaMA and Qwen2.5-Coder) employ a standard transformer block, where each feed-forward network (FFN) consists of: 
1) an \textit{up-projection} that expands hidden states,
2) a \textit{gating mechanism} that modulates feature activation, and
3) a \textit{down-projection} that compresses representations back to the model dimension.
In this section, we focus on UltraEdit based on its empirical performance observed in earlier RQs.
In UltraEdit, we compare three injection strategies: 
1) up-proj (default),
2) down-proj,
3) gate-proj, evaluated on LLaMA3.1-8B and Qwen2.5-Coder-7B under three decoding temperatures.


\begin{figure}[h]
    \centering
    \includegraphics[width=0.95\linewidth]{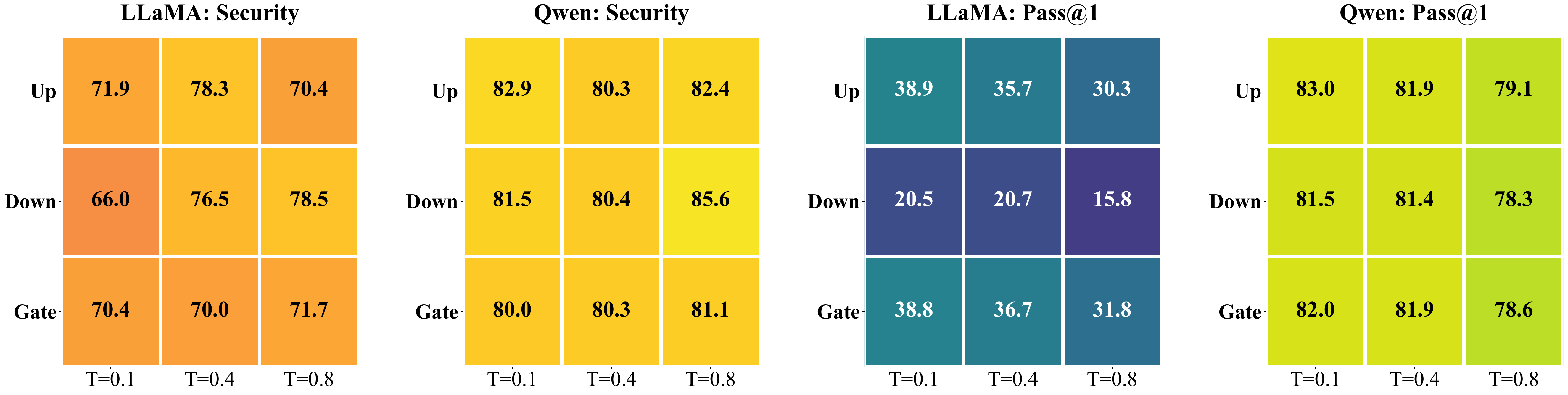}
    \vspace{-0.2cm}
    \caption{Impact of parameter injection location on security effectiveness and Pass@1.}
    \vspace{-0.05cm}
    \label{fig:rq7}
\end{figure}

Figure~\ref{fig:rq7} shows that up-proj consistently yields the best security--correctness balance.
Down-proj causes the most severe functional degradation: on LLaMA3.1-8B at $T=0.1$, Pass@1 drops from 38.9\% to 20.5\%, as edits near the residual stream directly corrupt representations reused by subsequent layers.
Gate-proj falls between the two but still underperforms up-proj.
The advantage of up-proj is consistent across both models, though the gap narrows on Qwen2.5-Coder-7B.
Up-projection operates where features are expanded and re-composed, allowing security-relevant distinctions to be introduced without overwriting core representations.

\vspace{-0.2cm}
\subsection{Editing Dataset Size}

\begin{wraptable}{r}{0.52\textwidth}
\vspace{-0.2cm}
\centering
\caption{Impact of editing dataset size on LLaMA3.2-3B.}
\vspace{-0.3cm}
\label{tab:dataset_size}
\tiny
\setlength{\tabcolsep}{3pt}
\renewcommand{\arraystretch}{1.2}
\begin{tabular}{lcccccc}
\toprule
\multirow{2}{*}{\textbf{Method}} &
\multicolumn{3}{c}{\textbf{Security Ratio (\%)}} &
\multicolumn{3}{c}{\textbf{Pass@1 (\%)}} \\
\cmidrule(lr){2-4} \cmidrule(lr){5-7}
& $T=0.1$ & $T=0.4$ & $T=0.8$ & $T=0.1$ & $T=0.4$ & $T=0.8$ \\
\midrule
UltraEdit-50   & 64.0 & 62.5 & 63.8 & 26.8 & 26.0 & 23.8 \\
UltraEdit-full & \textbf{70.8} & \textbf{70.0} & \textbf{65.9} & \textbf{27.1} & \textbf{26.5} & \textbf{22.9} \\
\midrule
DINM-50        & 63.1 & 59.5 & 61.7 & \textbf{23.1} & \textbf{23.2} & \textbf{21.5} \\
DINM-full      & \textbf{70.6} & \textbf{76.2} & \textbf{77.2} & 12.8 & 10.2 & 6.3 \\
\bottomrule
\end{tabular}
\vspace{-0.1cm}
\end{wraptable}
We investigate the impact of editing dataset scale by comparing models edited on 50 samples (\texttt{-50}) versus the full dataset of 710 pairs (\texttt{-full}) on LLaMA3.2-3B, evaluating UltraEdit and DINM under three decoding temperatures $T \in \{0.1, 0.4, 0.8\}$.
Table~\ref{tab:dataset_size} shows that larger editing datasets consistently improve security ratios for both methods.
However, the two methods diverge sharply in how dataset scale affects functional correctness.
UltraEdit-full achieves stronger security while largely preserving Pass@1 (27.1 vs.\ 26.8 at $T=0.1$), indicating that its localized parameter updates can absorb more security knowledge without disrupting general coding representations.
In contrast, DINM-full achieves substantially higher SR across all temperatures, but at a severe functional cost: Pass@1 drops from 23.1 to 12.8 at $T=0.1$, suggesting that cumulative neuron-level updates increasingly interfere with representations shared by general code synthesis.


\vspace{-0.2cm}
\subsection{Validity of Static-Analysis-Based Security Evaluation}
\label{sec:discussion-static-analysis}

Our security evaluation uses CodeQL as a scalable and reproducible static-analysis oracle for CWE-specific vulnerability detection. 
This design allows us to compare a large number of generated programs across different hardening methods under the same evaluation protocol. However, we do not treat CodeQL as a complete semantic oracle or a formal proof of security. 
Like other static analyzers, CodeQL may produce false positives when it conservatively over-approximates risky patterns, and false negatives when a vulnerability requires more precise path-, value-, or runtime-sensitive reasoning.
Therefore, the reported security ratio should be interpreted as a static-analysis-based security proxy rather than an absolute guarantee of security.
To reduce this validity threat, we use a validated query setup rather than an ad-hoc set of checks.
Our evaluation uses 41 CodeQL rules, including 29 official queries and 12 scenario-specific queries inherited from prior secure code generation benchmarks~\cite{He2023,Li2024}. 
In the original benchmark, the queries were manually inspected, and one false positive and one false negative were reported to the CodeQL team and subsequently fixed.\footnote{\url{https://github.com/github/codeql/issues/12770}, \url{https://github.com/github/codeql/issues/12753}.} 
We directly use the corrected query set. 
We also manually inspected 40 generated outputs balanced between CodeQL-positive and CodeQL-negative cases. 
CodeQL agrees with manual inspection on 37 out of 40 cases, yielding 92.5\% accuracy, 95.0\% precision, and 90.5\% recall. 
The three disagreements include one false positive and two false negatives.

Figure~\ref{fig:examples} illustrates the two main disagreement patterns. 
In the false-positive case, CodeQL flags a risky formatting pattern, but the formatted value is bounded and the destination buffer is sufficiently large, making the reported overflow infeasible in this context.
In the false-negative case, the code allocates memory and immediately dereferences the returned pointer without checking whether allocation succeeds, leading to a potential \texttt{NULL} dereference that is missed by the applied query. 
These examples show that CodeQL is useful for detecting CWE-relevant insecure patterns, but its results still require careful interpretation.

\begin{figure}[h]
\noindent
\begin{minipage}[t]{0.49\textwidth}
\begin{mycode}{Listing 1a: FP example}
/* FP: infeasible overflow. */
char buf[100];
double x = ((double) rand()) / RAND_MAX;
sprintf(buf, "%.2f", x);
\end{mycode}
\end{minipage}
\hfill
\begin{minipage}[t]{0.49\textwidth}
\begin{mycode}{Listing 1b: FN example}
/* FN: missing allocation-failure check. */
person *p = (person *) malloc(sizeof(person));
p->status = 0;
\end{mycode}
\end{minipage}
\vspace{-0.3cm}
\caption{Simplified examples illustrating CodeQL false-positive and false-negative cases.}
\vspace{-0.2cm}
\label{fig:examples}
\end{figure}

We further complement the original CodeQL-based evaluation with CodeGuard+~\cite{Fu2024}, a recent secure code generation benchmark that uses both CodeQL and SonarQube for vulnerability detection. 
This complementary evaluation reduces the risk that our conclusions are an artifact of a single static-analysis tool. Finally, we clarify that we do not treat arbitrary output differences as faults.
Differences in variable names, code organization, library choices, or semantically equivalent implementations are considered benign variability.
We regard a difference as an actual fault only when it corresponds to a verifiable security or functional problem, such as triggering a relevant static-analysis rule or failing task-level tests.

\color{black}

\vspace{-0.3cm}
\section{Threats to Validity}
\textbf{Representativeness of Vulnerability Scenarios.}
Our experiments focus on high-impact CWE categories from widely used secure code generation datasets and prior benchmarks~\cite{Li2024, Li2025, He2023, He2024}, improving comparability with existing work and covering commonly studied vulnerability patterns~\cite{MITRECorporation2024, Hajipour2024, Pearce2023}. 
To reduce benchmark-specific bias, we include multiple CWE categories and separate \textit{seen} and \textit{unseen} settings, where $\mathcal{T}_{seen}\cap\mathcal{T}_{unseen}=\emptyset$. 
Nevertheless, no fixed benchmark can exhaustively cover real-world vulnerabilities, and our findings may not fully transfer to emerging or underrepresented CWE types.
\textbf{Validity of Evaluation Oracles and Metrics.}
Following prior studies~\cite{Li2024, Li2025, He2023}, we measure security using static-analysis-based vulnerability detection and functional correctness using execution-based tests. CodeQL provides a scalable and reproducible CWE-specific signal, but it is not a complete semantic oracle and may produce false positives or false negatives. We mitigate this threat with a validated query setup and manual inspection of 40 generated outputs, where CodeQL agrees with manual inspection on 37 cases. Thus, we interpret Security Ratio as a static-analysis-based proxy rather than an absolute security guarantee. We also emphasize that CodeQL is used only as an evaluation oracle, not as a defense: it detects vulnerabilities after generation, whereas model editing aims to prevent insecure patterns before generation. HumanEval is used to measure relative functional regressions, and we further include MultiPL-E and CodeGuard+ to complement the evaluation with multi-language functional tests and an additional static analyzer. Therefore, our conclusions rely on relative trends across methods rather than any single metric as a complete measure of security or correctness.
\textbf{Scope of Model Editing and Baselines.}
Our study focuses on open-weight models because model editing requires access to parameters, leaving API-only proprietary models outside our editing scope, where inference-time defenses such as CoSec remain more applicable. 
Thus, our results compare two complementary paradigms rather than replacing one with the other. 
\toolname{} is not conceptually restricted to UltraEdit: its edit-aware regularization can anchor any parameter subset modified by a base editing method and could in principle be applied to other editors such as DINM. 
We instantiate \toolname{} on UltraEdit because it achieves the best security--correctness trade-off among the evaluated editing methods; applying it to more aggressive editors such as DINM may require stronger regularization and is left for future work.
Although we evaluate multiple model families, temperatures, and random seeds, implementation choices may affect absolute results. 
We release the data, scripts, outputs, and manual validation records to support reproducibility.

\vspace{-0.3cm}
\section{Related Work}
\textbf{Security Hardening for LLM-Generated Code.}
Recent studies have shown that LLMs can generate insecure or vulnerable code under certain contexts, motivating a growing body of work on security-aware code generation and hardening.
Existing approaches include vulnerability-aware data curation, security-constrained decoding, and post-hoc filtering or repair.
Among them, CoSec~\cite{Li2024, Li2025} represents the state-of-the-art, introducing a supervised co-decoding framework that leverages an auxiliary security model to steer generation at inference time.
While effective, CoSec relies on additional models, incurs non-trivial inference overhead, and its performance depends on the capacity and generalization ability of the auxiliary model.
Beyond CoSec, GitHub Copilot~\cite{GitHubCopilot2022} incorporates an LLM-based vulnerability prevention mechanism that mimics the behavior of static analysis tools, enabling real-time detection and suppression of insecure coding patterns during code completion.
He and Vechev~\cite{He2023} propose SVEN, a prefix-tuning-based hardening approach that freezes the base model and injects security signals through learnable prefix modules, demonstrating improved security across a range of real-world applications.
\noindent\textbf{Model Editing for LLMs.}
Model editing modifies specific LLM behaviors or knowledge without full retraining.
Existing techniques and benchmarks~\cite{Meng2022, Yao2023, Zhang2024a, Zhu2020, Meng2022a, Li2024b} mainly target factual correction and knowledge updates in general-purpose LLMs.
In the code domain, Gu \etal~\cite{Gu2023} use neuron-level editing to correct next-token errors, and CLMEEval~\cite{Li2024a} evaluates editing on code generation and summarization.
However, these studies focus on syntactic or semantic correctness, leaving security-relevant behaviors and vulnerability mitigation largely unexplored.

Different from the above studies, our study presents the first systematic empirical study on applying model editing to secure code generation.
We focus on sequence-level secure code generation, and conduct a comprehensive comparison between model editing techniques and CoSec.
Beyond security effectiveness, we analyze generalization, functional correctness, robustness, and efficiency.

\vspace{-0.3cm}
\section{Conclusion and Future Work}
\label{sec:con}

This paper presents the first comprehensive empirical study of using \emph{model editing} to harden LLM-based code generation, with a systematic comparison to the state-of-the-art inference-time hardening method, CoSec.
We find that model editing often achieves stronger security on seen CWEs and avoids CoSec’s dual-model inference overhead, but it can also introduce functional regressions, revealing a clear security–correctness trade-off.
To further improve the best-performing editing baseline, UltraEdit, we propose \toolname{}, a post-edit refinement strategy that combines functional tuning with edit-aware regularization to mitigate functional regressions while preserving the injected security behavior.
In addition, our ablation results show that editing outcomes are highly sensitive to key design choices, including injection depth, parameter location, and suffix context length.
Future work will investigate hybrid hardening mechanisms that better balance security generalization and functional preservation.


\vspace{-0.3cm}
\section*{Data Availability}
All datasets and code used in this study are available in our replication package at \url{https://github.com/swf1996120/SafeEdit}.

\section*{Acknowledgments}
This research is supported by Natural Science Foundation of Jiangsu Province (BK20251458).
This research is also supported by the Ministry of Education, Singapore under its Academic Research Fund Tier 3 (Award ID: MOET32020-0004). Any opinions, findings and conclusions or recommendations expressed in this material are those of the author(s) and do not reflect the views of the Ministry of Education, Singapore.
We would also like to thank the anonymous reviewers for their valuable feedback and suggestions.

\bibliographystyle{ACM-Reference-Format}
\balance
\bibliography{sample-base}

\end{document}